\documentclass[11pt]{article}
\usepackage{graphicx}
\usepackage{epsfig}
\usepackage{amssymb}
\usepackage{cite}
\usepackage{subfigure}
\DeclareMathSymbol{\lesssim}      {\mathrel}{AMSa}{"2E}
\DeclareMathSymbol{\gtrsim}       {\mathrel}{AMSa}{"26}
\newcommand{\als}{\alpha_s}
\newcommand{\RE}{{\rm Re}}
\newcommand{\IM}{{\rm Im}}
\newcommand{\Li}{{\rm Li}}
\def\be{\begin{equation}}
\def\ee{\end{equation}}
\def\bc{\begin{center}}
\def\ec{\end{center}}
\def\bea{\begin{eqnarray}}
\def\eea{\end{eqnarray}}

\def\nn{\nonumber}
\def\h4w{h_{4 w}}
\def\q4w{q_{4 w}}

\def\as{\alpha_s}

\def\sq2{\sqrt{2}}
\def\sb{s_\beta}

\def\mususy{\mu_{\rm S}}
\newcommand{\muh}{\mu_H}
\newcommand{\gi}{{\scriptscriptstyle G}} %

\newcommand{\mt}{m_t}

\newcommand{\cb}{c_\beta}

\newcommand{\xgu}{x_1}
\newcommand{\xgd}{x_{2}}
\newcommand{\yuiu}{y_{\tilde{u}_1}^{i}}
\newcommand{\yuid}{y_{\tilde{u}_2}^{i}}
\newcommand{\ydiu}{y_{\tilde{d}_1}^{i}}
\newcommand{\ydid}{y_{\tilde{d}_2}^{i}}
\newcommand{\zu}{z_{1}^{i}}
\newcommand{\zd}{z_{2}^{i}}

\catcode`@=11
\def\marginnote#1{}
\newcount\hour
\newcount\minute
\newtoks\amorpm
\hour=\time\divide\hour by60
\minute=\time{\multiply\hour by60 \global\advance\minute by-\hour}
\edef\standardtime{{\ifnum\hour<12 \global\amorpm={am}%
        \else\global\amorpm={pm}\advance\hour by-12 \fi
        \ifnum\hour=0 \hour=12 \fi
        \number\hour:\ifnum\minute<10 0\fi\number\minute\the\amorpm}}
\edef\militarytime{\number\hour:\ifnum\minute<10 0\fi\number\minute}
\def\draftlabel#1{{\@bsphack\if@filesw {\let\thepage\relax
   \xdef\@gtempa{\write\@auxout{\string
      \newlabel{#1}{{\@currentlabel}{\thepage}}}}}\@gtempa
   \if@nobreak \ifvmode\nobreak\fi\fi\fi\@esphack}
        \gdef\@eqnlabel{#1}}
\def\@eqnlabel{}
\def\@vacuum{}
\def\draftmarginnote#1{\marginpar{\raggedright\scriptsize\tt#1}}
\def\draft{\oddsidemargin 0.0truein
        \def\@oddfoot{\sl preliminary draft \hfil
        \rm\thepage\hfil\sl\today\quad\militarytime}
        \let\@evenfoot\@oddfoot \overfullrule 3pt
        \let\label=\draftlabel
        \let\marginnote=\draftmarginnote
   \def\@eqnnum{(\theequation)\rlap{\kern\marginparsep\tt\@eqnlabel}%
\global\let\@eqnlabel\@vacuum}  }
\catcode`@=12
\newenvironment{appendletterA}
 {
  \typeout{ Starting Appendix \thesection }
  \setcounter{section}{0}
  \setcounter{equation}{0}
  \renewcommand{\theequation}{A\arabic{equation}}
 }{
  \typeout{Appendix done}
 }

\begin{document}

\thispagestyle{empty}
\begin{flushright}
  RM3-TH/05-11 \\
  Roma-1416/05 \\

\end{flushright}
\begin{center}
  \vspace{1.7cm}
  \bc
  {\LARGE\bf QCD corrections to the electric dipole moment}\\[6pt]
  {\LARGE\bf of the neutron in the MSSM}

  \ec
  \vspace{1.4cm}
  {\Large \sc Giuseppe Degrassi$^{a}$, Enrico Franco$^b$,}\\
  {\Large \sc Schedar
    Marchetti$^{a}$ and Luca Silvestrini$^{b}$}

  \vspace{1.2cm}

  ${}^a$
  {\em
    Dipartimento di Fisica, Universit\`a di Roma Tre, INFN, Sezione di 
    Roma III, \\
    Via della Vasca Navale~84, I-00146 Rome, Italy}
  \vspace{.3cm}

  ${}^b$
  {\em  INFN, Sezione di Roma, Dipartimento di Fisica, 
    Universit\`a di Roma ``La Sapienza'',\\}
  {\em P.le Aldo Moro 2, I-00185 Rome, Italy}

\end{center}

\vspace{0.8cm}

\centerline{\bf Abstract}
\vspace{2 mm}
\begin{quote}
  We consider the QCD corrections to the electric dipole moment of the
  neutron in the Minimal Supersymmetric Standard Model. We provide a master formula
  for the Wilson coefficients at the low energy scale
  including for the first time the
  mixing between the electric and chromoelectric operators and correcting
  widely used previous LO estimates. We show
  that, because of the mixing between the electric and chromoelectric 
  operators, the neutralino contribution is always strongly suppressed.
  We find that, in general, the effect of the QCD corrections is to reduce the
  amount of CP violation generated at the high scale. We discuss
  the perturbative uncertainties of the LO
  computation, which are particularly large for the gluino-mediated
  contribution. This motivates our Next-to-Leading order analysis. We
  compute for the first time the order $\alpha_s$ corrections to the
  Wilson coefficients for the gluino contributions, and recompute the two-loop
  anomalous dimension for the dipole operators. We show that the large LO
  uncertainty disappears once NLO corrections are taken into account.
\end{quote}

\vfill
\newpage
\setcounter{equation}{0}
\setcounter{footnote}{0}
\vskip2truecm
\section{Introduction}
\label{sec:intro}

CP violation plays a twofold role in SUSY model building. On the one hand,
it is one of the main motivations to invoke New Physics (NP), since
within the Standard Model (SM) it is not possible to construct a
successful theory of baryogenesis, and also in the Minimal
Supersymmetric Standard Model (MSSM) electroweak baryogenesis calls
for additional sources of CP violation beyond the SM single phase in
the Cabibbo-Kobayashi-Maskawa (CKM) quark mixing matrix \cite{CW}. On
the other hand, CP violating processes provide very stringent
constraints on NP.  Indeed, the recent experimental progress in the
study of FCNC processes allows us to conclude that most probably NP
cannot contribute substantially (i.e. more than $\sim 20-30 \%$) to flavour
and CP violation in $\Delta S=2$ and $\Delta B=2$ processes
\cite{UTfitNP}. Since the MSSM contains tens of new sources
of flavour and CP violation, this experimental observation is quite
puzzling, leading to the so-called SUSY flavour problem, which is one
of the main open issues in SUSY model building~\cite{Pellicani}.

The new sources of CP violation present in the MSSM can be divided in
two groups: the first one contains new phases that appear in flavour
conserving quantities, the second contains those new sources of CP
violation that are also new sources of flavour violation. While the
latter are strongly constrained by $K$ and $B$ physics, at least for
those flavour-changing parameters that connect the first generation to
the other two,\footnote{To get a feeling for numbers, the imaginary
  parts of squark mass terms connecting the first two generations are
  constrained to be $\lesssim 10^{-3}-10^{-5}$ of the average squark
  mass.} the new sources of CP violation that are not directly
connected to flavour violation have little impact on FCNC processes
\cite{masieroporod} and are mainly constrained by the Electric Dipole
Moments (EDM) of the electron and of the neutron
\cite{ibrahim,Pokorski,altri}.  A careful analysis of EDM processes is
therefore mandatory in order to assess the allowed size of these NP
contributions to CP violation and their possible effects in
electroweak baryogenesis and in other CP violating processes.

Surprisingly enough, while the study of FCNC processes in the MSSM has
recently witnessed considerable theoretical advances, with the
inclusion of Next-to-Leading Order (NLO) QCD corrections \cite{SUSYNLO} and with the
computation of the relevant hadronic matrix elements with Lattice QCD
\cite{SUSYBPAR}, not only no corresponding effort has been made in the
study of the EDM of the neutron, but even incorrect LO results have
been widely used in the literature.

The present work aims to be a first step towards  bringing the EDM analysis in
SUSY at the same level of accuracy as the other FCNC and CP violation
studies. In particular, we focus on the perturbative aspects of the
QCD corrections to the neutron EDM.

The paper is organized as follows. In section \ref{sec:LO}
we discuss the neutron EDM at the LO in QCD: we give a
complete LO formula for the EDM, correcting some errors in previous
analyses, and we discuss the interplay of the various contributions.
We then study the uncertainties related to the LO approximation, and we find
that they are particularly large for gluino contributions. This motivates us
to upgrade the analysis to NLO that is presented in the next section.
We introduce the NLO QCD evolution
and the NLO matching conditions for the gluino contribution. We show that
after the inclusion of the NLO contributions
the scale uncertainty is reduced down to $\sim 2.5\%$. Finally we present
some conclusions.

\section{Leading Order Analysis}
\label{sec:LO}

In this Section, we provide the full LO expressions for the neutron
EDM in the MSSM, we discuss the interplay of the various SUSY
contributions and we study the uncertainties of the LO approximation.

\subsection{Anomalous dimension}
\label{sec:anomdim}
We write the relevant CP-violating effective low-energy Hamiltonian as
\be
H_{CPV} = \sum_q C_1^q (\mu)  O_1^q (\mu)   +
          \sum_q C_2^q (\mu)  O_2^q (\mu) +
          C_3 (\mu) O_3 (\mu)~~,
\label{eq:heff}
\ee
where
\begin{eqnarray}
O_1^q &=& -\frac{i}2 e \,Q_q  m_q \,\bar{q} \sigma^{\mu \nu} \gamma_5 q
             \, F_{\mu \nu}, \nn \\
O_2^q &=& -\frac{i}2 g_s m_q \bar{q}\, \sigma^{\mu \nu} t^a \gamma_5 q
           \, G^a_{\mu \nu} , \nn \\
O_3 &=& -\frac16 g_s f^{a b c} G^a_{\mu \rho} G^{b \rho}_{\nu}
  G^c_{\lambda \sigma} \epsilon^{\mu \nu \lambda \sigma}\,.
  \label{eq:oldops} 
\end{eqnarray}
The index $q$ runs over light quarks, and $Q_q=(2/3,-1/3)$ for up- and
down-type quarks respectively.
With this choice, all the operators have dimension six.

Defining $\vec{C} = (C_1^q,C_2^q,C_3)$ we write the renormalization group
equation for the Wilson coefficients as:
\be
\frac{d\, \vec{C}(\mu)}{d\, \ln \mu} = \gamma^T  \vec{C}(\mu)
\label{eq:RGE}
\ee
where at the LO $\gamma (\als) = (\als /4 \pi)\, \gamma^{(0)}$.

Let us now discuss the LO anomalous dimensions of the operators in
eq.~(\ref{eq:oldops}). The anomalous dimensions of operators $O_1^q$
and $O_2^q$ can be easily gleaned from that of the operators $O_7$ and
$O_8$ relevant in the $b \to s \gamma$ process (see ref.~\cite{ciu}).
The anomalous dimension of the Weinberg operator $O_3$ \cite{op:wein}
and of its mixing with
$O_2$ was derived in ref.~\cite{braaten}.  Therefore we get the
following LO anomalous dimension matrix:
\begin{equation}
  \label{eq:dimanom12}
  \gamma^{(0)} \equiv
\left(
    \begin{array}{ccc}
      \gamma_e & 0 & 0 \\
      &\\
      \gamma_{qe} & \gamma_q & 0 \\
      & \\
      0     & \gamma_{\gi q}       & \gamma_\gi
    \end{array}
  \right) =
\left(
    \begin{array}{ccc}
      8 \,C_F & 0 & 0 \\
      &\\
      8\, C_F & 16\, C_F - 4 \,N  & 0 \\
      & \\
      0     & - 2 \, N                  & N + 2 n_f + \beta_0
    \end{array}
  \right)
\end{equation}
where $C_F = 4/3,\, N= 3, \, \beta_0 = \frac{1}3(11 N - 2 n_f)$ with $n_f$
the number of active flavours.
The Wilson coefficients at the hadronic scale $\muh$  can be
easily obtained from those at a high scale $\mususy$  from
\bea
C_1^q (\muh)&=&   {\eta}^{\kappa_{e}} C_1^q(\mususy) +
          \frac{ \gamma_{qe} }{  \gamma_{e} - \gamma_{q}}
     \left( {\eta}^{ \kappa_{e}}-
   {\eta}^{ \kappa_{q}} \right) \, C_2^q(\mususy) + \nn \\
&&   \left[
\frac{\gamma_{\gi q} \gamma_{qe} \,{\eta}^{\kappa_{e}}}
       {(\gamma_{q} - \gamma_{e})( \gamma_{\gi} -\gamma_{e})}
+ \frac{\gamma_{\gi q} \gamma_{qe} \,{\eta}^{\kappa_{q}}}
       {(\gamma_{e}-\gamma_{q})( \gamma_{\gi} -\gamma_{q})}
+ \frac{\gamma_{\gi q} \gamma_{qe} \,\eta^{\kappa_{\gi}}}
  {(\gamma_{e}-\gamma_{\gi})( \gamma_{q} -\gamma_{\gi})}
\right] \, C_3(\mususy)~,  \label{C1} \\
C_2^q (\muh)&=&   {\eta}^{\kappa_{q}} C_2^q(\mususy) +
          \frac{ \gamma_{\gi q} }{  \gamma_{q} - \gamma_{\gi}}
     \left( {\eta}^{ \kappa_{q}}-
   {\eta}^{ \kappa_{\gi}} \right) \, C_3(\mususy)~,  \label{C2} \\
C_3 (\muh)&=&   {\eta}^{\kappa_{\gi}} C_3(\mususy)~, \label{C3}
\eea
where $ \eta = \als (\mususy)/\als(\muh)$ and
$\kappa_i = \gamma_i/(2 \beta_0)$.

The operator basis in eq.~(\ref{eq:oldops}) is very suitable to discuss
the anomalous dimension matrix. However, in order to avoid the explicit
appearance of the strong coupling at  the low scale in the operators,
it is more convenient to introduce a slightly different operator basis
\begin{eqnarray}
 O_e^q &=& -\frac{i}2 e \,Q_q  m_q \,\bar{q} \sigma^{\mu \nu} \gamma_5 q
             \, F_{\mu \nu}, \nn \\
  O_c^q &=& -\frac{i}2  m_q \bar{q}\, \sigma^{\mu \nu} t^a \gamma_5 q
           \, G^a_{\mu \nu} , \nn \\
  O_\gi &=& -\frac16 f^{a b c} G^a_{\mu \rho} G^{b \rho}_{\nu}
  G^c_{\lambda \sigma} \epsilon^{\mu \nu \lambda \sigma}
  \label{eq:ourops}
\end{eqnarray}
that defines our electric dipole ($O_e$), chromoelectric dipole ($O_c$) and
Weinberg operator ($O_\gi$) and
whose corresponding Wilson coefficients can be easily obtained
from eqs.~(\ref{C1}--\ref{C3}) by
redefining  the coefficients as
follows:
\bea
g_s(\muh)\, C_2^q (\mususy) &=&  \eta^{-(1/2)} C_c^q (\mususy), \nn  \\
g_s(\muh)\,  C_3 (\mususy) &=& \eta^{-(1/2)} C_\gi (\mususy). \label{relC}
\eea
To illustrate in a simple way the relation between the Wilson coefficients at
the $\mususy$ scale  and those at the $\muh$ scale we take $\mususy \sim \mt$
and  assume five flavours of light quarks between the scales
$\mususy$ and $\muh$, obtaining
\bea
C_e^q (\muh)&=&   {\eta}^{{16 \over 23}} C_e^q (\mususy) +
   8  \left( {\eta}^{{16 \over 23}}-
   {\eta}^{{14 \over 23}} \right)  \, { C_c^q (\mususy) \over g_s(\mususy)}
 \mathbf{+}
\frac{24}{85} \left[ 17 \, {\eta}^{{16 \over 23}}- 15\, {\eta}^{{14 \over 23}}
- 2 \,{\eta}^{{31 \over 23}} \right] \, {C_\gi (\mususy) \over g_s(\mususy)},
\label{Ce} \\
C_c^q (\muh)&=&   {\eta}^{{5 \over 46}} C_c^q (\mususy) +
          \frac{9 }{ 17}
     \left( {\eta}^{{5 \over 46}}-
   {\eta}^{ {39 \over 46}} \right) \, C_\gi (\mususy),  \label{Cc}\\
C_\gi (\muh)&=&   {\eta}^{{39 \over 46}} C_\gi (\mususy). \label{Cg}
\eea
It is interesting to note that in eqs.~(\ref{Ce}--\ref{Cg}) all the $\eta$'s
are raised to a positive power and then act as  suppression factors.

\begin{table}[t]
  \centering
  \begin{tabular}{|c|c|c|c|c|c|}
    \hline
    $X_{11}^{23} $&$ 1.25857 $&
    $X_{12}^{12} $&$ -9.78321 $&
    $X_{12}^{23} $&$ 10.06853 $\\
    $X_{13}^{12} $&$ -4.63415 $&
    $X_{13}^{23} $&$ 5.33040 $&
    $X_{13}^{35} $&$ -1.60476 $\\
    $X_{13}^{43} $&$ 7.76606 $&
    $X_{13}^{52} $&$ -7.05911 $&
    $X_{22}^{11} $&$ 1.22290 $\\
    $X_{23}^{11} $&$ 0.57927 $&
    $X_{23}^{34} $&$ -1.30387 $&
    $X_{23}^{51} $&$ 0.88239 $\\
    $X_{33}^{34} $&$ 2.17311 $& & & & \\
    \hline
  \end{tabular}
  \caption{Magic numbers $X_{ij}^{ab}$ for the evolution from six to four
flavours. See the text for details.}
  \label{tab:magic}
\end{table}

In general, SUSY masses are expected to be above $m_t$ while the
hadronic matrix  element is evaluated at a scale of the order of the
neutron mass.
In this  situation it is more appropriate to consider the evolution
from $\mususy > m_t$ to $\muh < m_b$, i.e. from the six- to the four-flavour
theory, that can be summarized via the so-called ``magic numbers''.
In this case, the  low-energy coefficients
$\vec{\rm C}(\muh) \equiv (C_e^q, C_c^q, C\gi)$ are given in terms of the high
energy  ones as
\begin{equation}
  \label{eq:magic64}
 {\rm C}_i (\muh) =
\sum_{j=1}^3 \sum_{a,b=1}^5 X_{ij}^{ab} \alpha_s(\mususy)^{Y_a}
  \eta^{Z_b} g_s(\mususy)^{\delta_{i1}(\delta_{j1}-1)} {\rm C}_j(\mususy)\,,
\end{equation}
with $Y_a$ and $Z_b$ given by:
\begin{eqnarray}
  Y_a
  &=&\left\{\frac{8}{75},\frac{64}{525},\frac{72}{175},\frac{163}{175},
    \frac{177}{175}\right\},\nonumber \\
  Z_b &=& \left\{\frac{3}{50}, \frac{14}{25}, \frac{16}{25},
    \frac{33}{50}, \frac{29}{25} \right\}
  \label{eq:magicnumbers}
\end{eqnarray}
and the nonvanishing entries in $X_{ij}^{ab}$ are listed in Table
\ref{tab:magic}.
These magic numbers have been obtained using the average
values $\overline m_t(m_t)=168.5$ GeV, $\overline m_b(m_b)=4.28$
GeV and $\alpha_s(M_Z)=0.119$.

A comparison with previous evaluations of the LO anomalous dimension
matrix is now in order. Several partial LO results are present in
the literature \cite{wise,dine,braaten,ALN} although the work of
ref.~\cite{ALN}, to be called ALN, can be regarded as the standard reference
for the QCD correction to the neutron EDM with its
numerical estimates of the QCD correction factors that have been and
are still widely used. With respect to ALN our analysis differs in two
aspects: i) we have included the mixing between the operators $O_e$
and $O_c$ that is neglected in ALN.  ii) Our definition of the
operator basis (eq.~(\ref{eq:ourops})) is different from that employed
in ALN. In particular, we write explicitly in the definition of the
operators $O_e$ and $O_c$ the mass of the quark, as well as in $O_e$
the charge of quark, while in the operator basis of ALN the quark mass
and charge is not present. Correspondingly the anomalous dimension
matrix of ALN should differ from ours by a factor $\gamma_m = -6\,
C_F$.  Taking into account this difference we find agreement with ALN
in the anomalous dimension result for the chromoelectric and Weinberg
operators. Instead, for the electric dipole operator we find that,
with the conventions employed by ALN, the anomalous dimension should
read $\gamma_e= -8/3$ , i.e. it has the opposite sign with respect to
the one quoted in ref.~\cite{ALN}.  As a consequence, the QCD
renormalization factor of the dipole operator, $\eta^{\rm ED}$, that
is estimated in ALN to be $\eta^{\rm ED} = 1.53$, should be $\eta^{\rm
  ED} < 1$ , i.e it does not enhance the CP violating effect but
actually suppresses it.  Employing the same values for strong coupling at
the high and low scale used in ALN we get $\eta^{\rm ED} \sim 0.61$.

In our view the definition of the operator basis we employ
(eq.~(\ref{eq:ourops})) has the advantage to make more transparent the
behavior of the perturbative QCD corrections to the neutron EDM that
in general give correction factors that decrease the amount of CP
violation generated at the high scale. In the ALN operator basis this
effect is somewhat hidden by the fact that the quark mass entering
their Wilson coefficients has to be taken at the high scale and
$m_q(\mususy) < m_q(\muh)$.  It should be noticed that the dependence
of the Wilson coefficients upon the quark mass can also appear in an
indirect way, e.g. through the matrices that diagonalize squark
masses.

\subsection{Hadronic Matrix Elements}
In order to compute the EDM of the neutron the matrix elements of
$ O_e^q,\, O_c^q,\, O_\gi$ between neutron states are also needed. At the
moment a result from Lattice QCD is not yet
available, although first steps in this direction have been recently
made~\cite{latticeEDM}. Several alternative approaches have been used
to estimate these matrix elements, as QCD sum rules \cite{qcdsum}
or chiral Lagrangians \cite{chiral}.   In this paper we are mainly concerned
about perturbative aspects of the EDM calculation, thus we are
going to use the simplest estimates of the operator matrix elements.
In particular, for the electric dipole operator, we use
the chiral quark model where the neutron is seen as a
collection of three valence quarks described by an $SU(6)$ symmetric
spin-flavour wave function.  In this model the neutron EDM is related
to that of the valence quarks by
\be
d^e_n= \frac{1}{3} \left( 4 d^e_d - d^e_u \right),
\label{eq:dn}
\ee
where
\be
 d^e_q =  e \, Q_q\, m_q(\muh)\, C^q_e(\muh)
  \label{eq:mee}
\ee
is the quark EDM. Concerning the contribution of the chromoelectric
and Weinberg operators to $d_n$ the simplest  estimate is obtained via
naive dimensional analysis \cite{georgi} giving
\bea
 d^c_n &= &  \frac{e}{4 \pi}\,\left( m_u(\muh)\,  C^c_u(\muh) +
             m_d(\muh)\,  C^c_d(\muh)\right), \label{eq:mec} \\
 d^\gi_n &= &  \frac{e}{4 \pi}\, \Lambda \, C_\gi (\muh)
  \label{eq:meg}
\eea
where $\Lambda \approx 1.19$ GeV is the chiral-symmetry-breaking scale.
We notice that in eqs.~(\ref{eq:mee}) and (\ref{eq:mec}) $m_q$ is
computed at the hadronic scale, while in the expressions for the
Wilson coefficients at the $\mususy$ scale the masses, as well
as $g_s$, are computed at the high scale.

\subsection{Wilson Coefficients in the MSSM}
\label{sec:wilcoeff}
The discussion in the previous sections has been general and applies
to any model in which $CP$ violating effects are generated at some
high scale. In this section we focus on the minimal supersymmetric
standard model (MSSM) with complex parameters assuming also that the
trilinear SUSY-breaking scalar couplings are proportional to the
corresponding Yukawa coupling. In this model besides the SM $CP$
violating phases $(\delta_{CKM}, \, \theta_{QCD})$ that will be
neglected in the present discussion as well as the flavour mixing,
there are new phases associated to the $\mu$-term in the
superpotential, the supersymmetry-breaking parameters of the gaugino
mass, the trilinear scalar couplings $A$, and the bilinear scalar term
in the Higgs potential.  However not all the phases of these
quantities are physical. It is then possible to assume the gaugino
masses and the bilinear term to be real so that we are left only with
two $CP$ violating phases, one associated with the $\mu$ term,
$(\phi_\mu)$, and the other with the trilinear scalar coupling,
$(\phi_A)$, that is in general flavour-dependent.  These phases will be
present in the mass matrices of squarks, charginos and neutralinos
inducing an EDM at the quark level. In particular in the squark mass
matrix
\be M^2_{\tilde{q}}=
\left(\begin{array}{cc} m_{\tilde{q}_{ L}}^2\, & m_q X_q \\
    m_q X_q^\ast\, &\,m_{\tilde{q}_R}^2\end{array} \right)
\label{squamm}
\ee
the only complex parameter is  the left-right mixing term
$X_u \equiv A^\ast_u - \mu \cot \beta,\: X_d \equiv A^\ast_d - \mu \tan \beta$,
where both phases are present. In $X_q$,
$\tan \beta = v_2/v_1$ is the ratio of the VEVs of the Higgs fields.
Instead in the chargino mass matrix
\be
M_{\tilde{C}}=\left(\begin{array}{cc} M_2 & \sqrt{2}\,s_\beta\,M_W\\
                        \sqrt{2}\,c_\beta\,M_W & \mu\end{array}\right)
\label{chamm}
\ee
as well as in the neutralino one
\be
M_{\tilde{N}}=\left(\begin{array}{cccc}
 M_1 & 0 & -c_{\beta}\,s_W\,M_Z & s_\beta\,s_W\,M_Z\\
   0 & M_2 & c_{\beta}\,c_W\,M_Z & - s_{\beta}\,c_W\,M_Z\\
 -c_\beta\,s_W\,M_Z & c_\beta\,c_W\,M_Z&0&-\mu\\
s_\beta\,s_W\,M_Z & -s_\beta\,c_W\,M_Z&-\mu&0\end{array} \right)
\label{neumm}
\ee
the only complex term is $\mu$ and therefore only $\phi_\mu$ is
present. In eqs.~(\ref{chamm},\ref{neumm}) $s_\beta \equiv \sin
\beta,\, c_\beta \equiv \cos \beta$, $s_W$, $c_W$ are the sine and
cosine of the weak mixing angle and $M_{1,2}$ are the soft-breaking
gaugino masses associated to the $U(1)$ and $SU(2)$ groups.

To present our results in a transparent way we perform the computation
of the Wilson coefficients using current eigenstates for squark
fields. In this basis the squark propagator is a $2 \times 2$ matrix
given by
\be
\Delta_{\tilde{q}} (k)=\frac{i}{(k^2-m_{\tilde{q}_1}^2)\,
(k^2-m_{\tilde{q}_2}^2)}\,
\left(\begin{array}{cc} k^2-m_{\tilde{q}_R}^2 & m_q X_q^\ast \\
 m_q X_q & k^2-m_{\tilde{q}_L}^2\end{array}\right)
\ee
where $m_{\tilde{q}_i}$ are the eigenvalues of $M^2_{\tilde{q}}$. Neglecting
${\cal O}(m_q^2)$ terms we have that $m_{\tilde{q}_1} \simeq m_{\tilde{q}_L},
\, m_{\tilde{q}_2} \simeq m_{\tilde{q}_R}$ so that $ \Delta_{\tilde{q}} (k)$
reduces to
\be
\Delta_{\tilde{q}} (k) \simeq \,i\,\left(\begin{array}{cc}
\frac{1}{k^2-m_{\tilde{q}_1}^2}&
\frac{ m_q X_q^\ast}{(k^2-m_{\tilde{q}_1}^2)\,(k^2-m_{\tilde{q}_2}^2)}\\
\frac{ m_q X_q }{(k^2-m_{\tilde{q}_1}^2)\,(k^2-m_{\tilde{q}_2}^2)}&
\frac{1}{k^2-m_{\tilde{q}_2}^2}
\end{array}\right).
\ee
We notice that within this approximation the left-right propagator is still
exact.

The Wilson coefficients of the operators $O_e, \, O_c$ are generated at the
one-loop order while that of $O_\gi$ appears for the first time at the
two-loop level. At the LO we can then set $C_\gi (\mususy) =0$ and write
\be
C_e^q (\mususy) =
C_{e \tilde{g}}^q (\mususy) +  C_{e \tilde{\chi}^-}^q (\mususy)
+  C_{e \tilde{\chi}^0}^q (\mususy)
\label{eq:cegcn}
\ee
and similarly for $C_c^q$.

We find for the gluino contribution
\bea
 C_{e \tilde{g}}^q (\mususy) &= & \frac{\alpha_{s}}{4\pi\, m_{\tilde{g}}^2}
\IM \left( \frac{ X_q}{m_{\tilde{g}}} \right) {8 \over 3}
   \,\tilde{B}\left(\xgu, \xgd \right),
\label{eq:gluinoe} \\
C_{c \tilde{g}}^q  (\mususy) &=&
\frac{g_s\, \alpha_{s}}{4\pi\, m_{\tilde{g}}^2}
\IM \left(\frac{X_q}{ m_{\tilde{g}}} \right)
   \,\tilde{C}\left(\xgu,\xgd\right)
\label{eq:gluinoc}
\eea
where $x_i =m_{\tilde{g}}^2/m_{\tilde{q}_i}^2$. The
explicit expressions for the functions
$\tilde{B}$ and $\tilde{C}$ as well as those entering in the chargino and
neutralino  contributions are  collected in the Appendix.

For the chargino we have
\bea
 C_{e \tilde{\chi}^-}^u (\mususy) &=&
\frac{\alpha\, Q_{u}^{-1}}{4\pi s^2_{W}}\,
\sum_{i=1}^2 \frac{1}{\sqrt{2}\,\sb\,\, m_{\tilde{\chi_{i}}^-}^2}\,
\left(\frac{\, m_{\tilde{\chi_{i}}^-}}{M_{W}}\right)\,\left\{
\IM \left(
V_{i2}\, U_{i1} \right)
\left[(Q_d-Q_u)\, A(\ydiu)+ Q_{d}\,
 B(\ydiu)\right]
\right. \nn \\
&& \left.  ~~~~-\IM \left( \frac{m_d X_d^\ast}{m_{\tilde{\chi_{i}}^-}^2}
  V_{i2}\, U_{i2} \right) Y_d
\left [ (Q_d-Q_u)\,
  \tilde{A}(\ydiu, \ydid)+
Q_d \,\tilde{B}(\ydiu, \ydid)\right] \right\},
\label{eq:chargeu}\\
 C_{e \tilde{\chi}^-}^d (\mususy) &=&
\frac{\alpha\, Q_{d}^{-1}}{4\pi s^2_{W}}\,
\sum_{i=1}^2 \frac{1}{\sqrt{2}\,\cb\,\, m_{\tilde{\chi_{i}}^-}^2}\,
\left(\frac{\, m_{\tilde{\chi_{i}}^-}}{M_{W}}\right)\,\left\{
\IM \left(
V_{i1}\, U_{i2} \right)
\left[(Q_u-Q_d)\,A(\yuiu)+ Q_{u}\,B(\yuiu)\right] \right. \nn \\
&& \left. ~~~~-\IM \left( \frac{m_u X_u^\ast}{m_{\tilde{\chi_{i}}^-}^2}
  V_{i2}\, U_{i2} \right) Y_u
\left [(Q_u-Q_d)\, \tilde{A}(\yuiu,\yuid)+
Q_u \,\tilde{B}(\yuiu,\yuid)\right] \right\},
\label{eq:charged}\\
 C_{c \tilde{\chi}^-}^u (\mususy) &=&
\frac{g_s\,\alpha}{4\pi s^2_{W}}\,
\sum_{i=1}^2 \frac{1}{\sqrt{2}\,\sb\,\, m_{\tilde{\chi_{i}}^-}^2}\,
\left(\frac{\, m_{\tilde{\chi_{i}}^-}}{M_{W}}\right)\,\left\{
\IM \left(
V_{i2}\, U_{i1} \right) B(\ydiu) \right. \nn \\
&& \left.
~~~~-\IM \left( \frac{m_d X_d^\ast}{m_{\tilde{\chi_{i}}^-}^2}
  V_{i2}\, U_{i2} \right) Y_d  \tilde{B}(\ydiu,\ydid)\right\},
\label{eq:chargcu}\\
 C_{c \tilde{\chi}^-}^d (\mususy) &=&
\frac{g_s\,\alpha}{4\pi s^2_{W}}\,
\sum_{i=1}^2 \frac{1}{\sqrt{2}\,\cb\,\, m_{\tilde{\chi_{i}}^-}^2}\,
\left(\frac{\, m_{\tilde{\chi_{i}}^-}}{M_{W}}\right)\,\left\{
\IM \left(
V_{i1}\, U_{i2} \right) B(\yuiu)  \right. \nn \\
&& \left.
~~~~-\IM \left( \frac{m_u X_u^\ast}{m_{\tilde{\chi_{i}}^-}^2}
  V_{i2}\, U_{i2} \right) Y_u \tilde{B}(\yuiu,\yuid)\right\}
\label{eq:chargcd}
\eea
where $y_{\tilde{q}_j}^{i}=
m_{\tilde{\chi}^-_i}^2/m_{\tilde{q}_j}^2$,
$U$ and $V$ are the matrices that diagonalize $M_{\tilde{C}}$ according
to $U^\ast M_{\tilde{C}} V^{-1} = M_{\tilde{C}}^D$
and $Y_{u,d}$ are the Yukawa couplings of the up and down quarks in units
of $e/s_W$.
In eqs.~(\ref{eq:chargeu}--\ref{eq:chargcd}) we have also written explicitly
the contributions proportional to the mass and the Yukawa coupling
of the light quarks to show that the phase combination that enters
in  the gluino contribution is actually present in the chargino term
in a suppressed way. Indeed, in the chargino contribution the only relevant
phase  is $\phi_\mu$, hidden inside the matrices $U$ and $V$. This can
be seen explicitly in the following simplified expression, obtained by
neglecting the contributions proportional to quark masses and Yukawa
couplings:
\begin{eqnarray}
C_{e \tilde{\chi}^-}^u (\mususy) &=&
\frac{\alpha\, Q_{u}^{-1}}{4\pi s^2_{W}}\,
\sum_{i=1}^2 \frac{1}{\sqrt{2}\,\sb\,\, m_{\tilde{\chi_{i}}^-}^2}\,
\left\{
\IM \left( \frac{ U_{i2}^*\, U_{i1} \mu}{M_W} \right)
\left[(Q_d-Q_u)\, A(\ydiu)+
Q_{d}\,B(\ydiu)\right]
\right\} \,.\nn\\ & &
  \label{eq:chargsimpl}
\end{eqnarray}
Analogous expressions can be obtained from
eqs.~(\ref{eq:charged})-(\ref{eq:chargcd}) with the substitutions:
\begin{eqnarray}
  m_{\tilde{\chi_{i}}^-}\, \IM \left(V_{i2}\, U_{i1} \right) &\to&
  \IM \left(U_{i2}^*\, U_{i1} \mu \right)\,, \nonumber \\
  m_{\tilde{\chi_{i}}^-}\, \IM \left(V_{i1}\, U_{i2} \right) &\to&
  \IM \left(V_{i2}^*\, V_{i1} \mu \right)\,.
  \label{eq:subsc}
\end{eqnarray}

Finally the neutralino contribution, neglecting terms proportional to the
the quark masses,  is given by:
\bea
 C_{e \tilde{\chi}^0}^q (\mususy) &=&
\frac{\alpha}{4\pi s^2_{W}}\,
\sum_{i=1}^4 \frac{1}{\, m_{\tilde{\chi_{i}}^0}^2}\,
\left\{
\IM \left(\frac{K^{a}_{qi} K^b_{qi} X_q}{m_{\tilde{\chi}_{i}^0}} \right)
\tilde{B}(\zu,\zd)\right.
\nonumber \\
&& \left. + \left(\frac{m_{\tilde{\chi_{i}}^0}}{M_{W}}\right)\,
\left(\IM \left( K^{b}_{qi} K^{c}_{qi} \right)
B(\zd) -
\IM \left( K^{a}_{qi} K^{c}_{qi} \right)
B(\zu)\right) \right\},
\label{eq:neue}\\
 C_{c \tilde{\chi}^0}^q (\mususy) &=& g_s\, C_{e \tilde{\chi}^0}^q
\label{eq:neuc}
\eea
where $z_j^{i}= m_{\tilde{\chi}^0_i}^2/m_{\tilde{q}_j}^2$,
and
\bea
K^{a}_{ui}&=&
  \sqrt{2}\left[ \left(Q_u-\frac12 \right) \tan \theta_{W}\,Z^{i1}+
  \frac12 Z^{i2}\right]\,,
\label{eq:Ku1} \\
K^{b}_{ui} &=& \sqrt{2}\tan \theta_{W}\,Q_{u}\,Z^{i1}\,,
\label{eq:Ku2}\\
K^{c}_{ui} &=& \frac{1}{\sqrt{2}\sb}Z^{i4}\,,
\label{eq:Ku3}\\
K^{a}_{di}&=&
  \sqrt{2}\left[ \left(Q_d+\frac12 \right) \tan \theta_{W}\,Z^{i1}-
  \frac12 Z^{i2}\right]\,,
\label{eq:Kd1} \\
K^{b}_{di} &=& \sqrt{2}\tan \theta_{W}\,Q_{d}\,Z^{i1}\,,
\label{eq:Kd2}\\
K^{c}_{di} &=& \frac{1}{\sqrt{2}\cb}Z^{i3}\,.
\label{eq:Kd3}
\eea
In eqs.~(\ref{eq:Ku1}-\ref{eq:Kd3}) $Z$ is the matrix that diagonalizes
$M_{\tilde{N}}$ according to  $Z^\ast M_{\tilde{N}} Z^{-1} = M_{\tilde{N}}^D$.
As can be seen from eq.~(\ref{eq:neue}) in the neutralino contribution
both $\phi_\mu$, through the matrix $Z$, and the $X_q$ phase
combination are actually present.

We noticed that the results reported in
eqs.~(\ref{eq:gluinoe})-(\ref{eq:chargcd}) and
(\ref{eq:neue})-(\ref{eq:Kd3}) are fully in agreement  with those
in ref.~\cite{ibrahim,Pokorski} and represent the lowest order approximation.
\subsection{LO results}
In this section we investigate, at the LO, the effect of QCD
corrections on the neutron EDM, to assess whether they can
significantly reduce the  individual gluino, chargino and
neutralino contributions making the EDM constraint on SUSY
phases less severe.

To discuss in a simple way the effect of the QCD corrections and in
particular the importance of the mixing between $O_e$ and $O_c$ that
was neglected in previous analyses, we consider
eqs.~(\ref{Ce}-\ref{Cg}) assuming $\eta = 0.3$ and setting $C_\gi =0$.
Then
\bea
   C_e^q (\muh)&=& 0.43\, C_e^q (\mususy) -0.38 \,{ C_c^q
  (\mususy) \over g_s(\mususy)},
  \label{Cen} \\
C_c^q (\muh)&=&  0.88 \,C_c^q (\mususy),   \label{Ccn}\\
C_\gi (\muh)&=&  0 .
\eea
Thus, if $C_e^q (\mususy)\simeq C_c^q (\mususy)/g_s (\mususy)$ the resulting
$C_e^q (\muh)$ is strongly
suppressed. With our definition of the operators the above situation is
achieved when the gluon in a diagram contributing to $C_c^q$ is attached
to a squark of the same charge of the external quark. This case is
realized in the neutralino contribution (see  eq.~(\ref{eq:neuc})).
Indeed, we can estimate the neutralino contribution to
$d_n^e,\, d_n^c$, as given in eqs.~(\ref{eq:dn},\ref{eq:mec}),
by employing the Wilson coefficient at  the low scale evaluated via
eqs.~(\ref{Ce}-\ref{Cg}) with $g_s(\mususy)=1.22$. We get
\bea
d_n^{e,\tilde{\chi}^0} &\simeq& \frac{e}3 \left[ 4 \,m_d(\muh)
\left(-\frac13 0.05\right) C_{e \tilde{\chi}^0}^d (\mususy) -
m_u(\muh) \left(\frac23 0.05\right)  C_{e \tilde{\chi}^0}^u (\mususy) \right]
\nn\\
& \approx &  e \left[- m_d(\muh) \,0.02 \,
 C_{e \tilde{\chi}^0}^d (\mususy) - 0.01\, m_u(\muh)\,
C_{e \tilde{\chi}^0}^u (\mususy) \right]~~,\label{eq:prova1} \\
d_n^{c,\tilde{\chi}^0} &\simeq& 0.88 \frac{e}{4 \pi} \left[
m_u(\muh) \, g_s(\mususy)\,C_{e \tilde{\chi}^0}^u (\mususy) + m_d(\muh) \,
g_s(\mususy)\,C_{e \tilde{\chi}^0}^d (\mususy)\right]\nn\\
&\approx& e \, 0.08 \, \left[ m_u(\muh) \, C_{e \tilde{\chi}^0}^u (\mususy) +
m_d(\muh) \, C_{e \tilde{\chi}^0}^d (\mususy)\right].
\label{eq:prova2}
 \eea
Eqs.~(\ref{eq:prova1}-\ref{eq:prova2})  show that the individual quark EDMs
are strongly suppressed by the QCD corrections. A so large effect is
actually  specific to the neutralino contribution because of the simple
relation between
$C_{e \tilde{\chi}^0}^q$ and $C_{c \tilde{\chi}^0}^q$ (eq.~(\ref{eq:neuc})).
The general case is more complicated and the resulting effect depends
upon the relative sign between $C_{e}^q (\mususy)$ and $C_{c}^q (\mususy)$.

It is not our purpose here to perform a general analysis of EDM
constraints on SUSY models. Rather, our aim is to illustrate the
impact of QCD corrections on the computation of the EDM. To do so, we
study a specific point in the SUSY parameter space that we choose  with a
mass spectrum similar to that of the benchmark point 1a of the Snowmass Points
and Slopes
\footnote{We use the low-energy
  spectrum obtained from the tool \cite{Kraml} available on the Web at
  the URL: {\tt http://kraml.home.cern.ch/kraml/comparison/}} (SPS)
\cite{SLHA}.
 The SPS benchmark points  are actually defined
assuming real parameters, however we take the mass spectrum of point
1a as indicative, also in the case of complex parameters, of possible
mass values of a ``typical'' mSUGRA scenario with an intermediate
value of $\tan \beta$. We take $ m_{\tilde{g}}=585$ GeV, $M_{1} = 100$
GeV, $M_{2} = 190$ GeV, $m_{\tilde{u}_1} = 540$ GeV, $m_{\tilde{u}_2} =
525$ GeV, $m_{\tilde{d}_1} = 550$ GeV, $m_{\tilde{d}_2} = 520$ GeV,
$|\mu| = 355$ GeV, $\tan \beta =10$, $|A_d| = 855$ GeV, $|A_u| = 675$ GeV
and $\mususy= 465$ GeV.
To obtain $d_n$ we have
chosen the hadronic scale $\muh = 2$ GeV with $m_d(\muh) = 7$ MeV and
$m_u(\muh) = 3$ MeV.
\begin{figure}[t]
  \begin{center}
    \includegraphics[height=0.45\textwidth,angle=270]{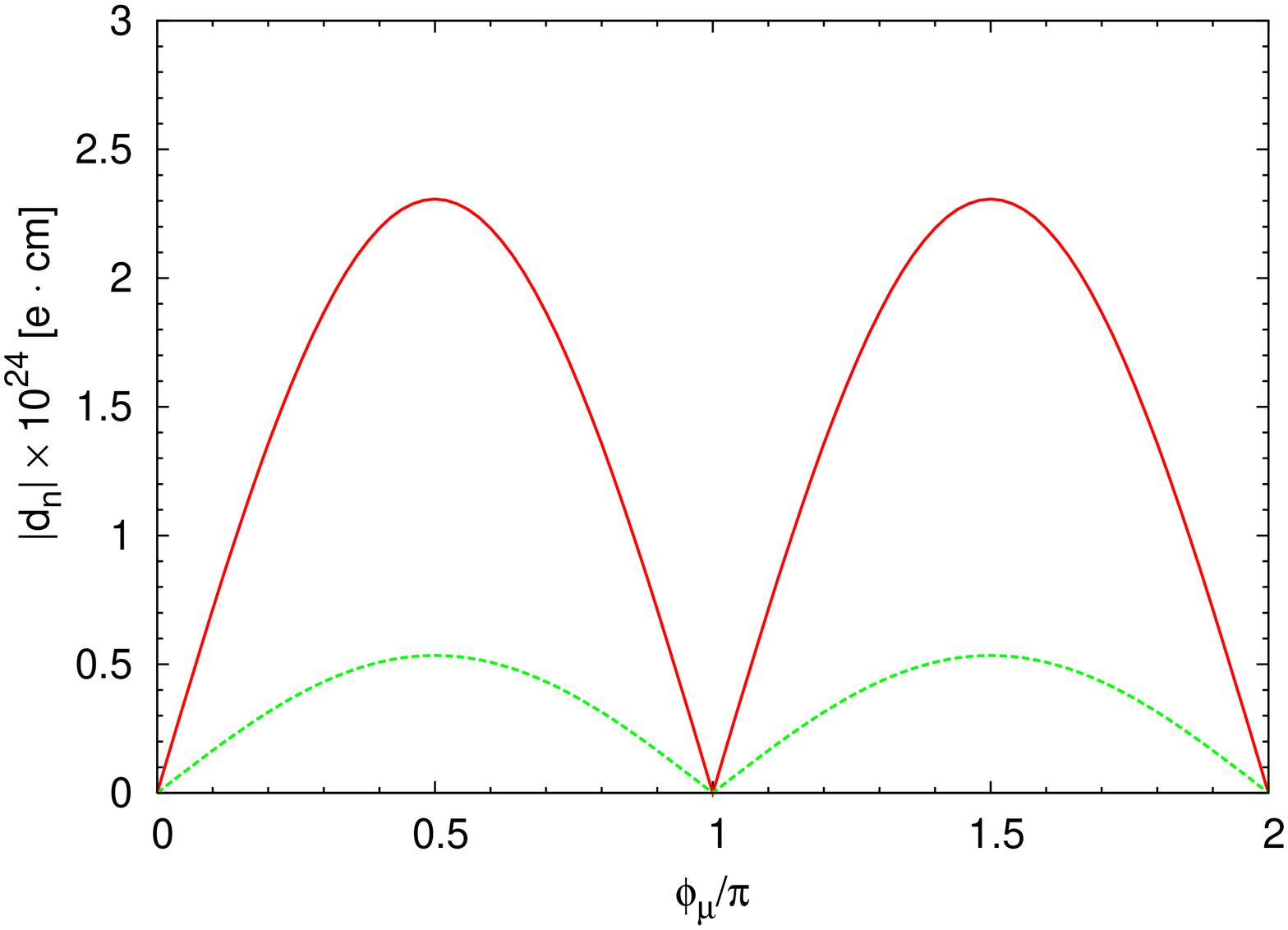}
    \includegraphics[height=0.45\textwidth,angle=270]{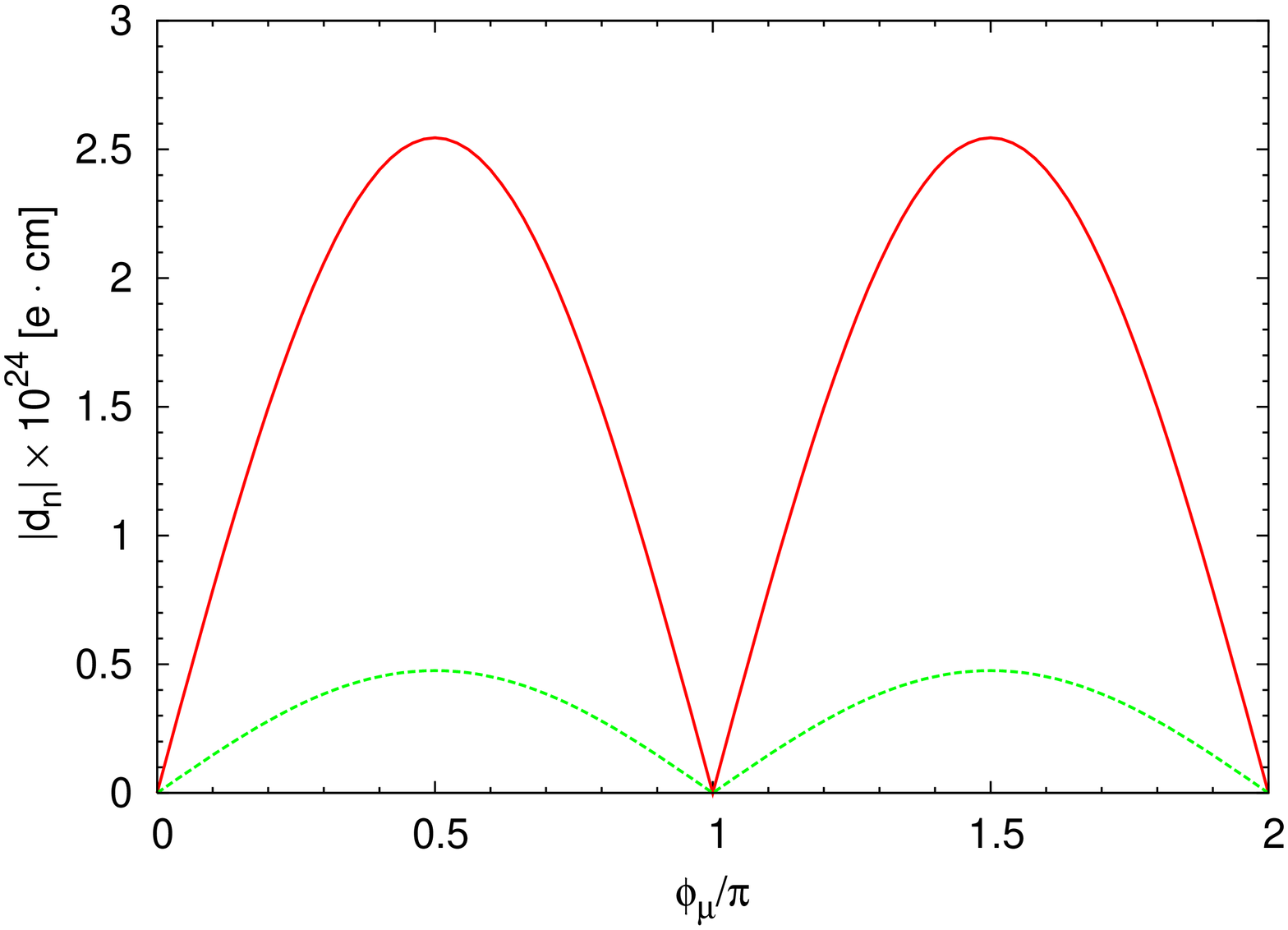}
    \includegraphics[height=0.45\textwidth,angle=270]{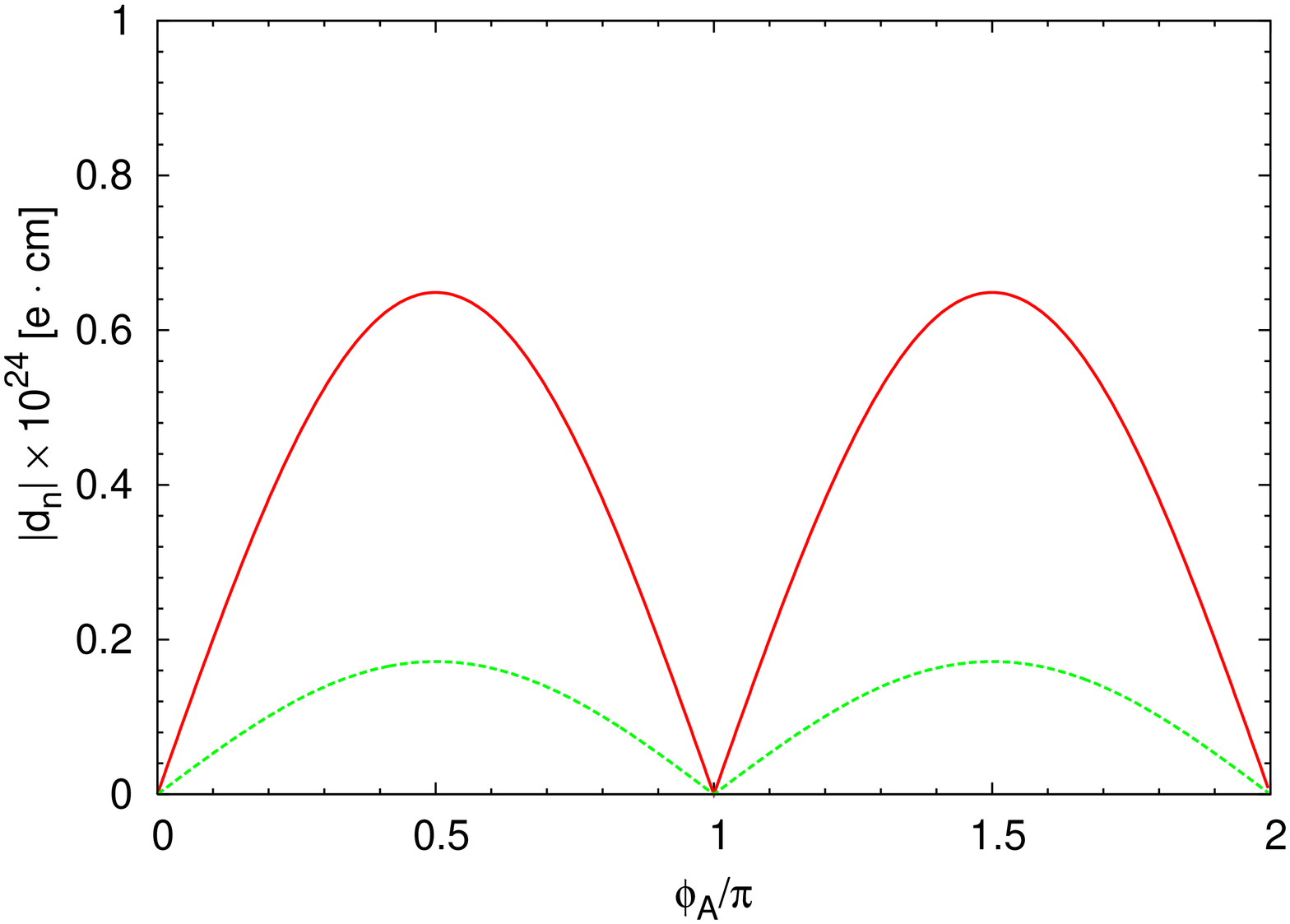}
    \includegraphics[height=0.45\textwidth,angle=270]{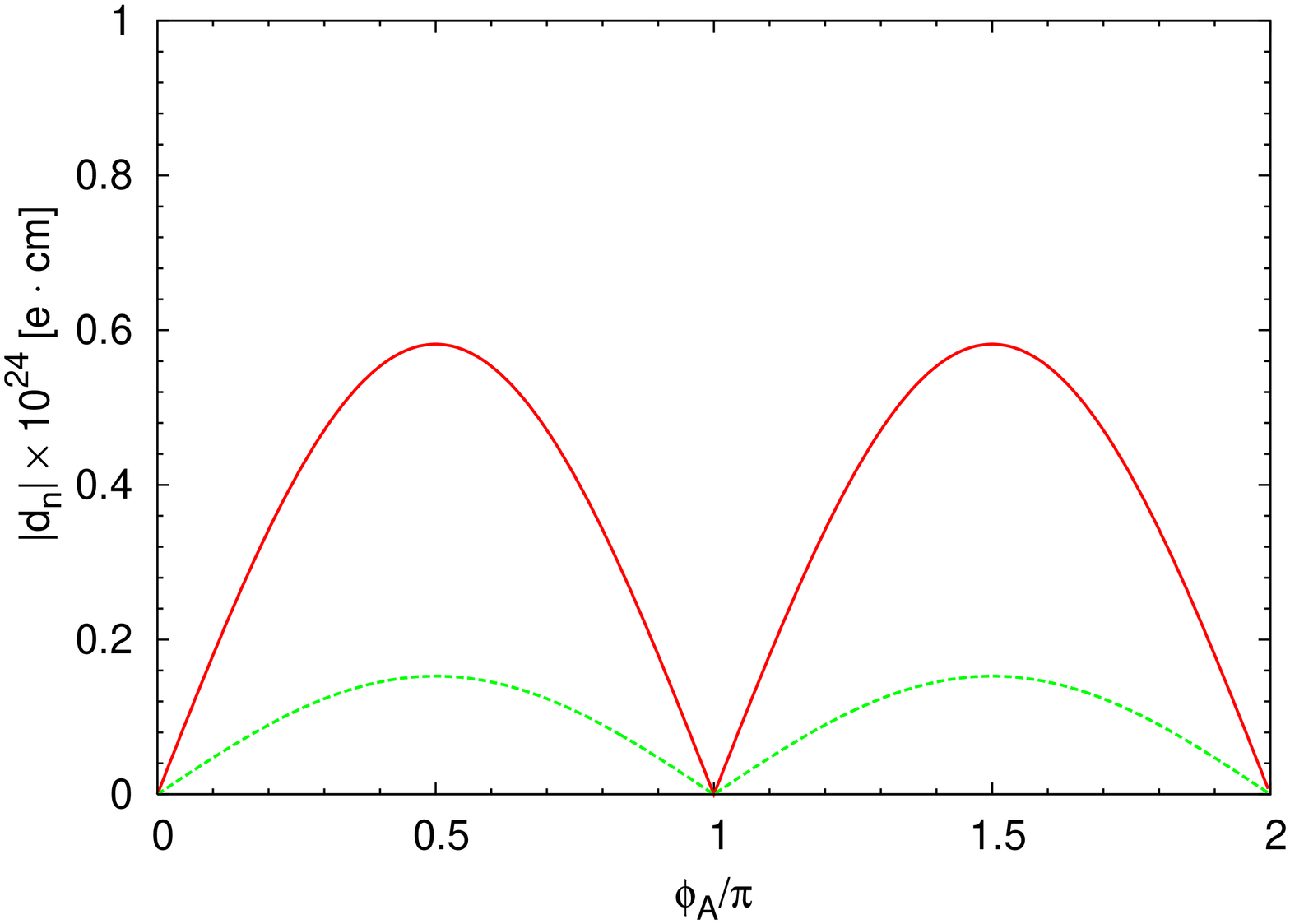}
    \caption{\itshape{Gluino contribution to the neutron EDM versus
        $\phi_\mu$ (top) and $\phi_{A_u}=\phi_{A_d}$ (bottom),
        without (left) and with (right) QCD corrections. The
        solid line is the $d_n^e$ contribution, while the dotted one
        the corresponding $d_n^c$ contribution.}}
    \label{fig:LOGlu}
  \end{center}
\end{figure}

In fig.~\ref{fig:LOGlu} we show the effect of QCD corrections on
the gluino contribution to the EDM of the neutron. In the figure we
plot the absolute value of $d_n$ as a function of $\phi_A$ and $\phi_\mu$
with the other SUSY parameters set to the values listed above.
\begin{figure}[h]
  \begin{center}
    \includegraphics[height=0.45\textwidth,angle=270]{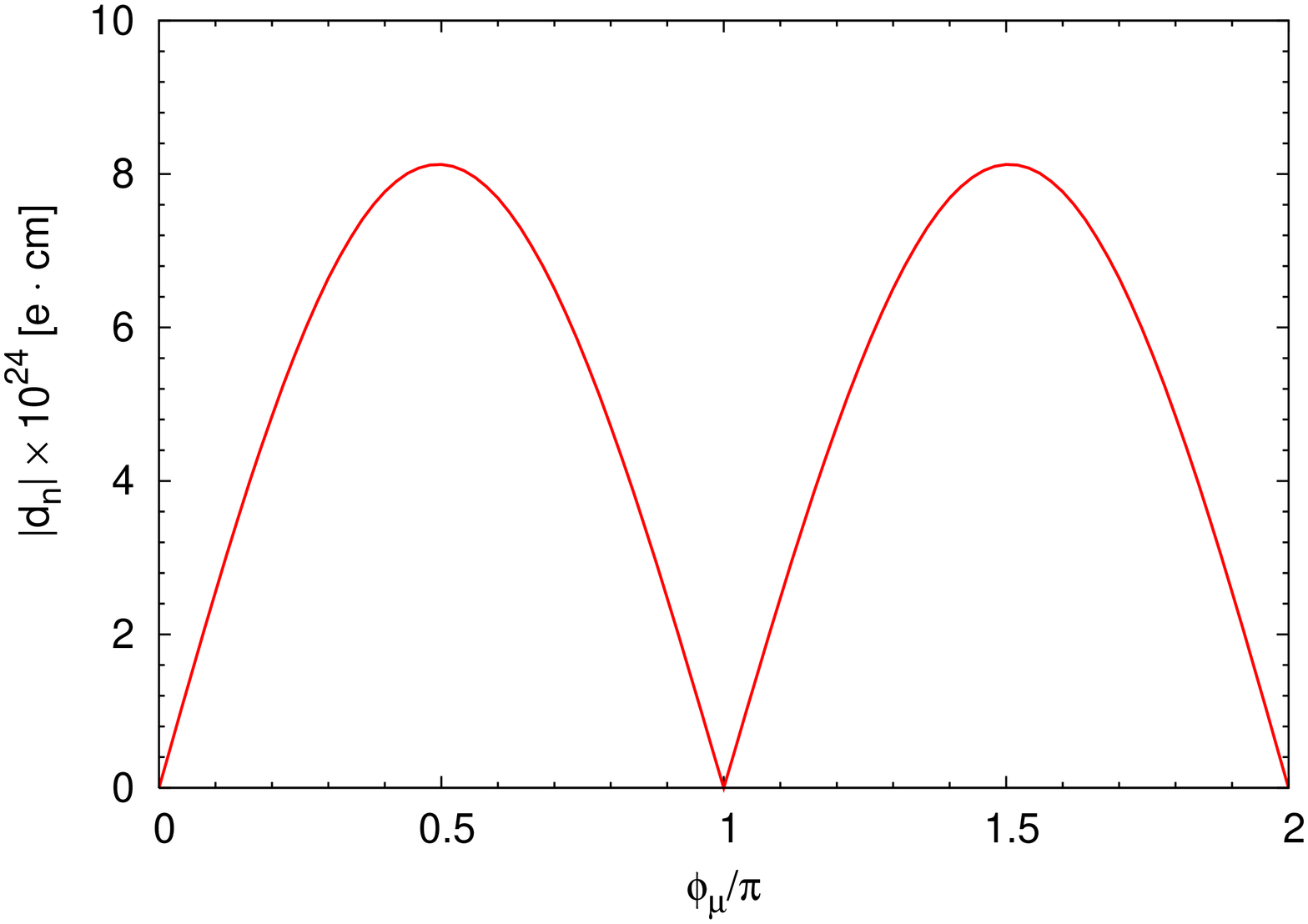}
    \includegraphics[height=0.45\textwidth,angle=270]{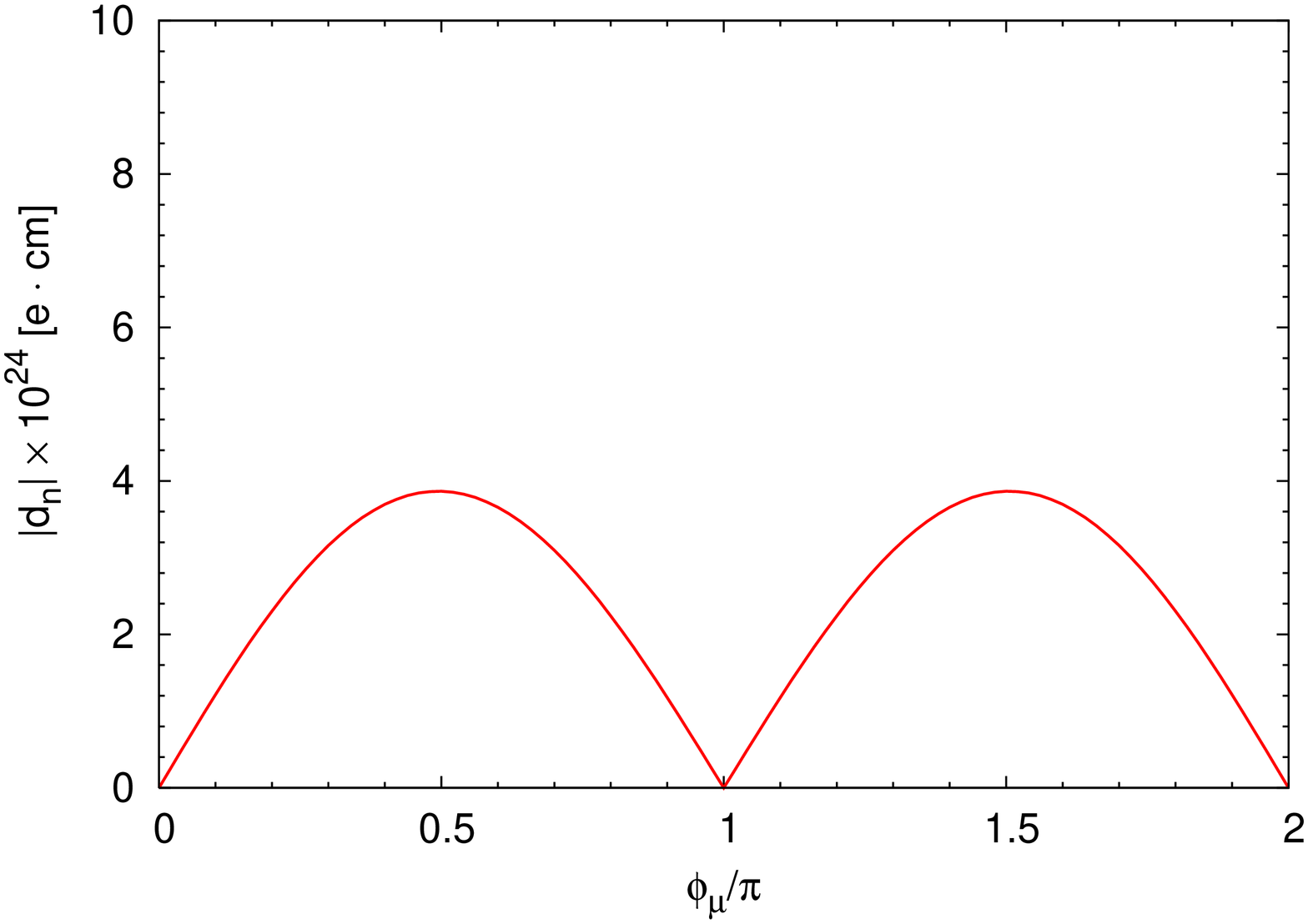}
    \caption{\itshape{Chargino contribution to the neutron EDM versus
        $\phi_\mu$, without (left) and with (right) QCD
        corrections. We show only the $d_n^e$ contribution because
        the $d_n^c$ one is negligible.}}
    \label{fig:LOCha}
  \end{center}
\end{figure}
\begin{figure}[h]
  \begin{center}
    \includegraphics[height=0.45\textwidth,angle=270]{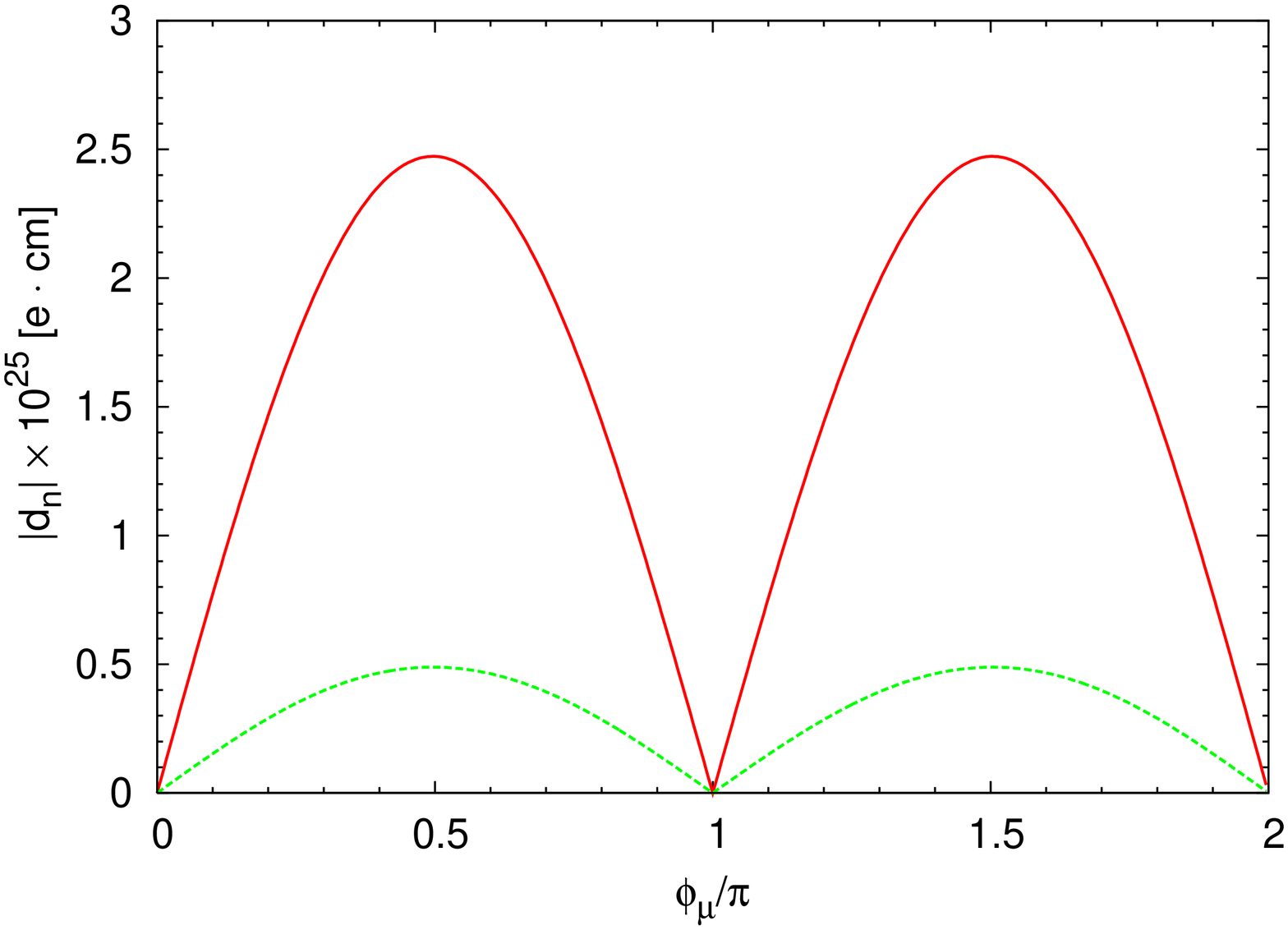}
    \includegraphics[height=0.45\textwidth,angle=270]{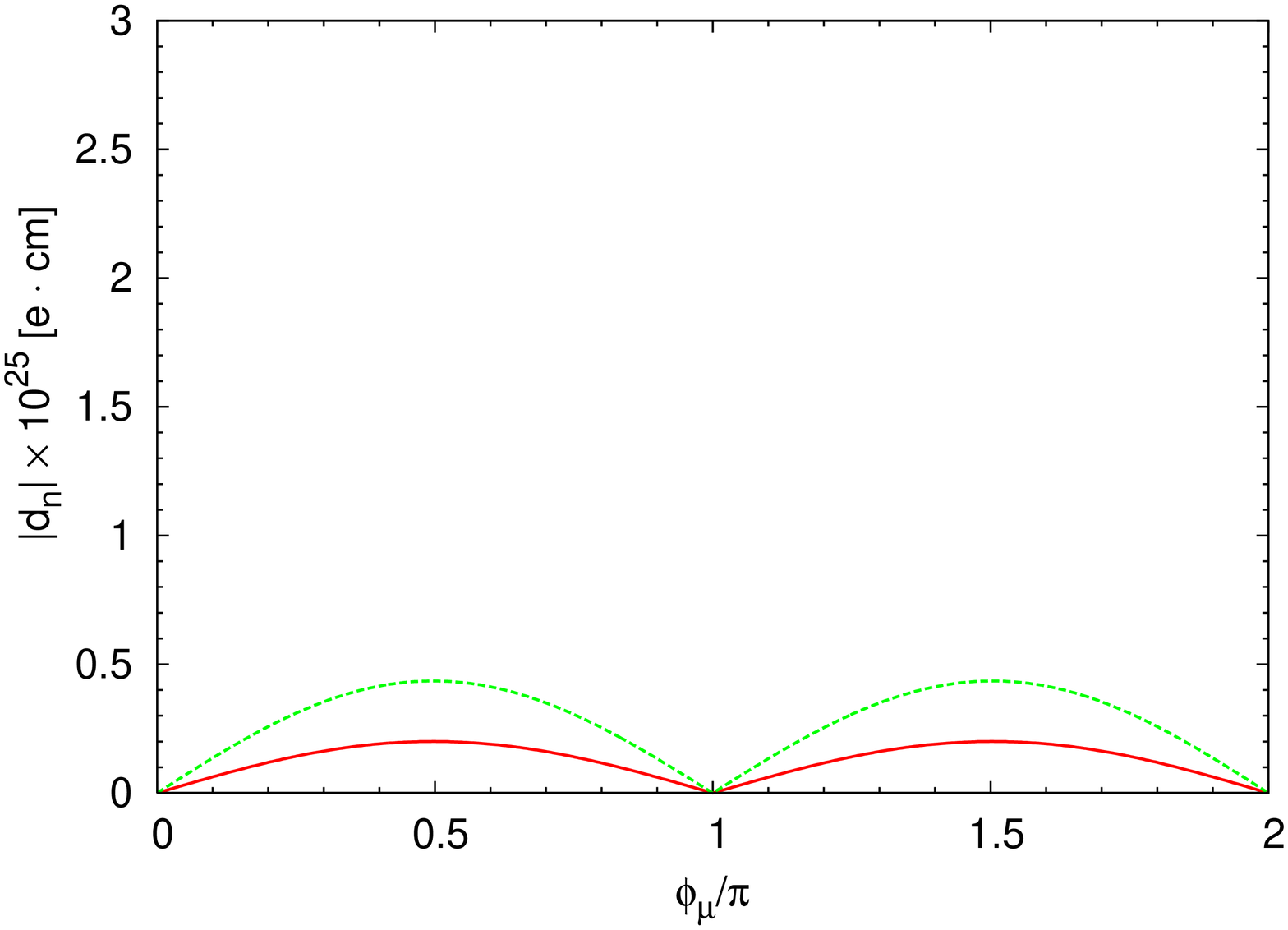}
    \includegraphics[height=0.45\textwidth,angle=270]{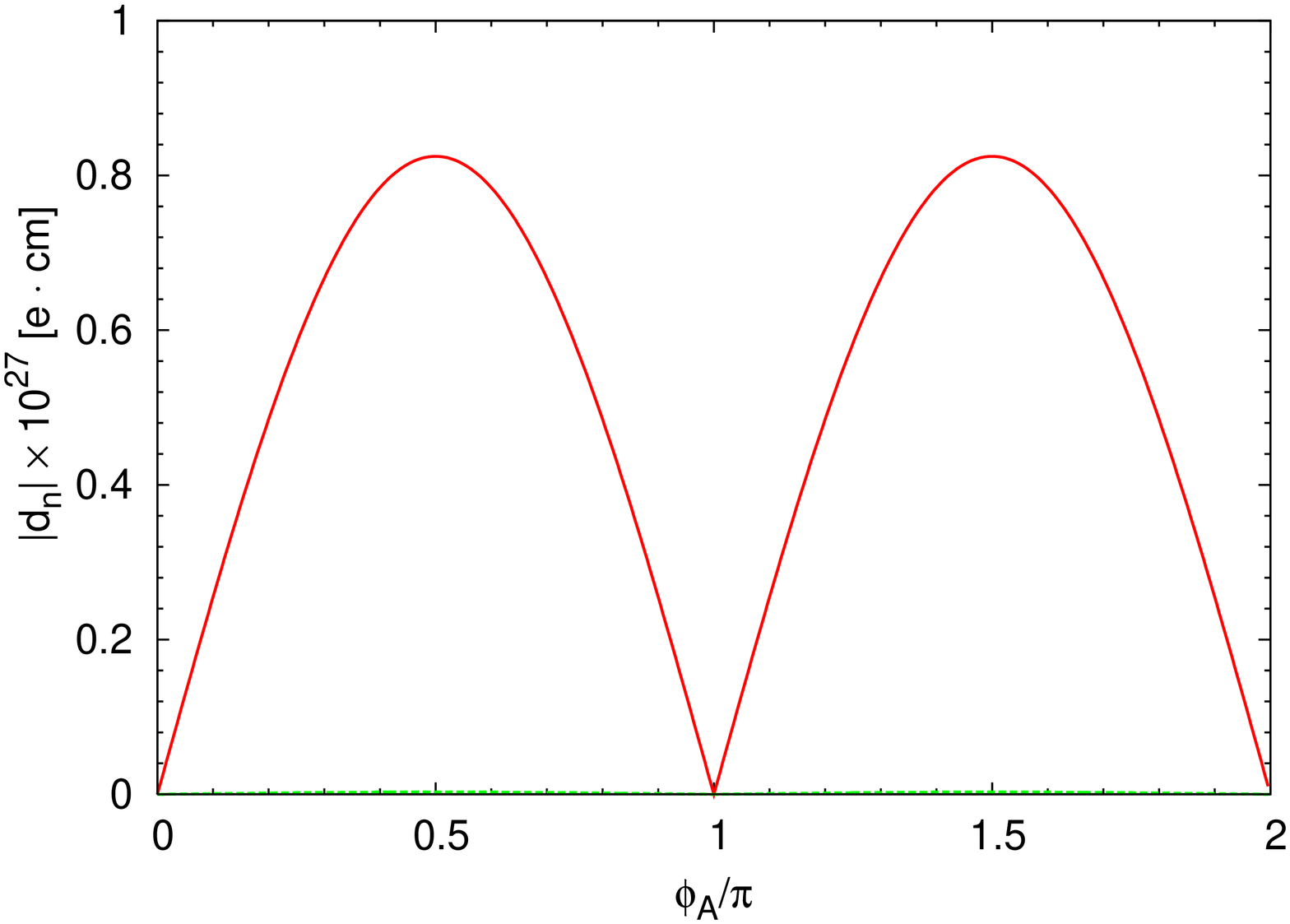}
    \includegraphics[height=0.45\textwidth,angle=270]{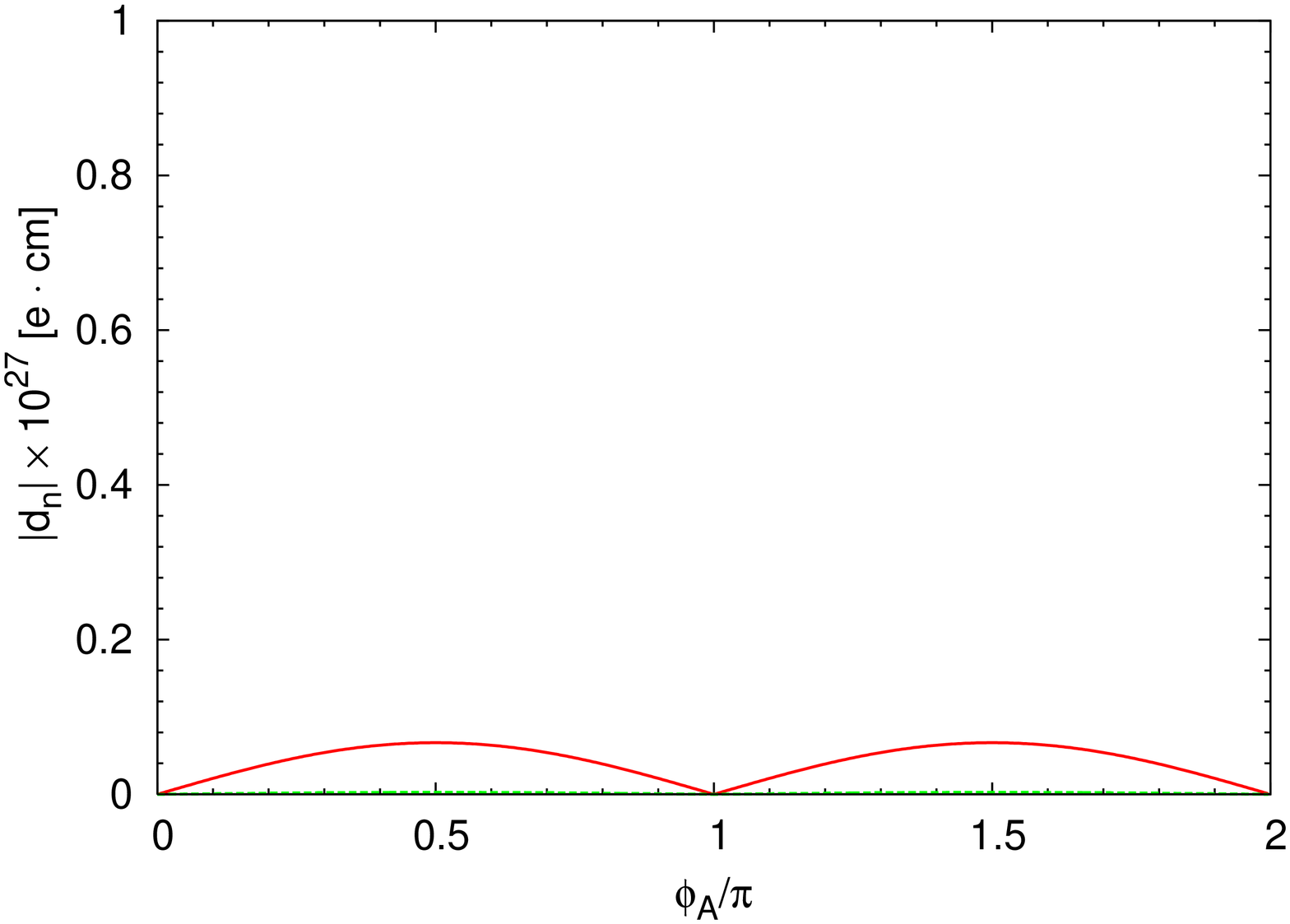}
    \caption{\itshape{ Neutralino contribution to the neutron EDM
        versus $\phi_\mu$ (top) and $\phi_{A_u}=\phi_{A_d}$ (bottom),
        without (left) or with (right) QCD corrections. The solid
        line is the $d_n^e$ contribution, while the dotted one the
        corresponding $d_n^c$ contribution.}}
    \label{fig:LONeu}
  \end{center}
\end{figure}
As can be seen
comparing the plots on the left, which are obtained without QCD
corrections, to the ones on the right, where QCD corrections are
included, the effect in this case amounts to $\sim 10\%$.
Fig.~\ref{fig:LOCha} and fig.~\ref{fig:LONeu} show the corresponding
analysis for chargino and neutralino contributions, respectively.
Fig.~\ref{fig:LOCha} shows that for chargino contributions the
inclusion of QCD corrections reduces the amount of CP violation
generated at the $\mususy$ scale by a factor $\sim 50\%$.
Finally, the simple analysis of the neutralino contribution discussed
above is substantiated by fig.~\ref{fig:LONeu} where this strong
reduction is clearly visible.
\clearpage

A popular mechanism \cite{ibrahim}
invoked to suppress the neutron EDM without resorting to extremely
small phases or very heavy SUSY particles is the search for regions of the
parameter space where cancellations among the three different contributions
are active. It is always possible to find regions
of the parameter space where contributions depending upon different
parameters cancel each other, although it can be questioned if
these regions can be representative of general situations.
With respect to this,
it is interesting to note that, since the neutralino contribution is
always much more suppressed by QCD corrections than the gluino and
chargino ones, the cancellation mechanism among different
contributions invoked in ref.~\cite{ibrahim} should actually work
between the gluino and chargino only. However, these two contributions
depend upon different phase combinations.  As an example, $\phi_A$ is
only present in the gluino contribution.

\subsection{Uncertainties of the LO analysis}
In the above analysis all the
uncertainties of the LO computation have been neglected. The uncertainties
connected to the nonperturbative evaluation of the hadronic matrix elements go
beyond the scope of this work, since we focus our analysis on the
perturbative aspects of QCD effects. Therefore, let us assume that
some nonperturbative method such as Lattice QCD will produce in the
future the necessary matrix elements at a scale $\muh=2$ GeV, so that
we fix the hadronic scale in our analysis. Then, we are left with the
uncertainties connected to the matching between the full and the
effective theory at the scale $\mususy$.

It is well known that in the RGE improved perturbation theory there
remain unphysical $\mususy$-dependences which are of the order of the
neglected higher order terms.  Usually, this uncertainty can be
estimated by varying the matching scale in a (arbitrarily chosen)
given range. However, for the EDM computation, there are further
sources of uncertainty. All contributions depend upon the squark masses, but
the precise definition of these masses cannot be fixed at LO, so that
one can use pole, $\overline{DR}$ or any other squark mass. Indeed,
the difference between the results obtained using two different mass
definitions is of higher order in $\alpha_s$ and provides an estimate
of this additional LO uncertainty. Furthermore, the gluino
contribution also depends on the gluino mass and, more important, on
the strong coupling. Neither the definition of the gluino mass nor,
in principle,  the scale of $\alpha_s$ is fixed at LO, so that they
constitute another source of uncertainty. All these uncertainties  can be
ameliorated only by a NLO calculation.
\label{sec:LOuncertainties}

\begin{figure}[ht]
  \begin{center}
    \includegraphics[height=0.45\textwidth,angle=270]{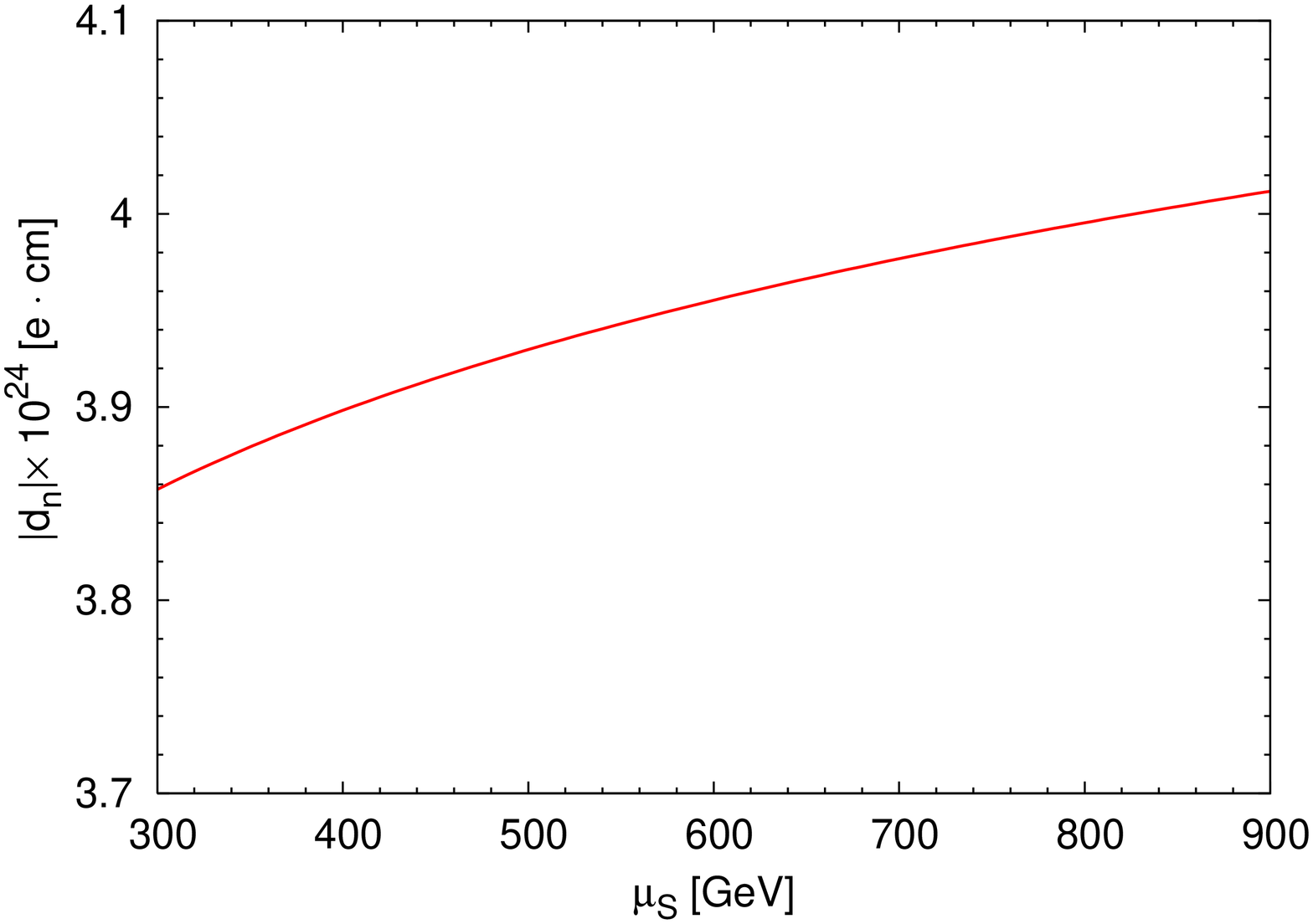}
   \includegraphics[height=0.45\textwidth,angle=270]{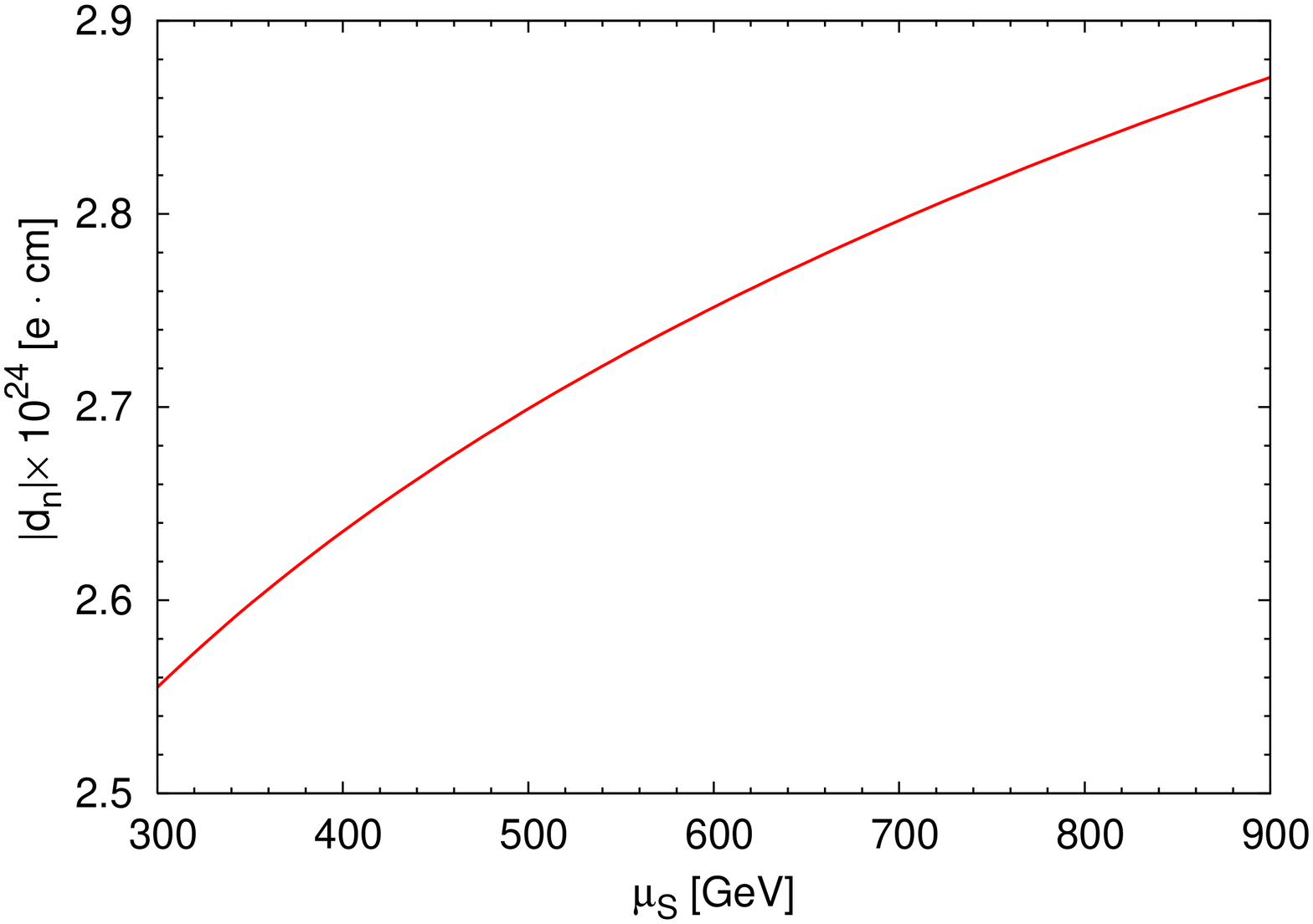}
    \caption{\itshape{The $\mususy$ scale dependence of the chargino
        (left) and gluino (right)
        contributions to $d_n^e$ at LO. }}
    \label{fig:scale}
  \end{center}
\end{figure}

In fig.~\ref{fig:scale} we illustrate the LO uncertainty only due to
the choice of the matching point. Here and in the following NLO
analysis we will use an average squark mass
$m_{\tilde{q}}^2=(m_{\tilde{q}_1}^2+m_{\tilde{q}_2}^2)/2$.  In the
figure we plot the LO gluino (right) and chargino (left) contributions
to $|d_n| \simeq |d_n^e|$ as a function of the matching scale with $\as$
and $\overline{DR}$ masses evaluated, for simplicity, at the scale
$\mususy$. As expected the $\mususy$ dependence is more pronounced in
the gluino case and amounts to 10-15 \% while in the chargino
contribution it reaches at most 4 \%.

The LO gluino contribution shows a  substantial uncertainty
that, including  all effects, can be expected to be $\sim 20\%$.
To reduce it to a level comparable to  that of the chargino contribution
one needs the NLO computation of this contribution that will be
discussed in the next section.

\section{Next-to-Leading  Order Analysis}
In this section we do not attempt to perform a complete NLO analysis of
the QCD corrections to the neutron EDM, instead  we focus on 
the relevant pieces needed to discuss the  reduction of the scale
dependence of the gluino contribution. 
We present the NLO anomalous dimension
matrix for the electric and chromoelectric operators and  the
NLO Wilson coefficients for the gluino contribution. For completeness we
present also the Wilson coefficients of the Weinberg operator.
We recall that at the LO
$C_\gi(\muh) =0$,
therefore to obtain the NLO result it is sufficient to
know the LO $\gamma_\gi,\, \gamma_{\gi q}$ entries of the anomalous dimension
matrix.
\subsection{NLO anomalous dimension}
The discussion of  the NLO anomalous dimension matrix is more easily
accomplished in the  $O_1^q$--$O_3$ basis of eq.(\ref{eq:oldops}).
Indeed in this basis the anomalous dimension matrix can be organized in powers
of $\as$ as
\be
\gamma = \frac{\als}{4\, \pi} \gamma^{(0)} +
         \left(\frac{\als}{4\, \pi}\right)^2  \gamma^{(1)} ,
\label{eq:nlogamma}
\ee
where $\gamma^{(0)}$ is given in eq.(\ref{eq:dimanom12})
and
\be
  \label{nloanomdim}
  \gamma^{(1)}   =
\left(
    \begin{array}{ccc}
      \left( {548 \over 9}N -16\, C_F - {56 \over 9} n_f \right) C_F & 0 & 0 \\
      & & \\
        \left( {404 \over 9}N -32 \,C_F - {56 \over 9} n_f \right) C_F   &
         -{458 \over 9}- {12 \over  N^2} +{214 \over 9}N^2 +
{56\, \over 9}\frac{n_f}N - {13  \over 9}N\,n_f & 0 \\
 & & \\
 $*$ & $*$ & $*$
    \end{array}
  \right) .
\ee
We have computed the NLO anomalous dimension in eq.(\ref{nloanomdim}) and
our results are in complete agreement with those obtained
from the known NLO anomalous dimension of the $O_7$ and $O_8$ operators in
the $b \to s \gamma$ process \cite{misiak}.

The entries in the third row are unknown but not needed at the NLO, since
they contribute only to the $(i3)$ sector of the NLO magic numbers, and the
initial coefficient of the Weinberg operator is vanishing  at the LO.
The corresponding Wilson coefficients in the $O_e^q$--$O_G$ basis can be easily
obtained using eq.(\ref{relC}).
\begin{table}[t]
  \centering
  \begin{tabular}{|c|c|c|c|c|c|}
    \hline
    ab & 43 & 43 & 54 & 55 & 63\\
    ij & 11 & 12 & 12 & 22 & 12\\
    \hline
$W_{ij}^{ab}$ & 11.301 & 85.158 & -79.353 & 9.9191 & 0 \\[1mm]
$R_{ij}^{ab}$ & -8.7762& -70.209 & 65.693 & -6.9887 & 0\\[1mm]
$\delta X_{ij}^{ab}$ & -0.39785 & -3.1828 & 3.7203 & -0.65874 & -0.46504 \\
    \hline
  \end{tabular}
  \caption{Magic numbers $R_{ij}^{ab}$ , $W_{ij}^{ab}$ and $\delta
    X_{ij}^{ab}$ for
   the NLO evolution from six to four flavours.}
  \label{tab:magicNLO}
\end{table}
The simplest way to present the NLO evolution of the Wilson coefficients is
via the magic numbers. Referring to the $O_e^q$--$O_G$ basis we can write for
a generic scale $\mu$
\be
  \vec {\rm C}(\mu) = \vec {\rm C}^{(0)}(\mu) +
\frac{\alpha_s(\mu)}{4\pi} \, \vec {\rm C}^{(1)}(\mu)
\label{eq:wilsonexp}
\ee
where ${\rm C}_i^{(0)}(\mususy)$ is the LO Wilson coefficient at the scale
$\mususy$ and its  evolution from $\mususy > m_t$ to $\muh < m_b$ is
given by eq.(\ref{eq:magic64}).
The evolution of $\vec {\rm C}^{(1)} (\mu)$ from
$\mususy > m_t$ to $\muh < m_b$ can  be summarized in the following
way
\begin{eqnarray}
  {\rm C}_i^{(1)} (\muh) &=&
 \eta \sum_{j=1}^3 \sum_{a=1}^6 \sum_{b=1}^5 \alpha_s(\mususy)^{Y_a}
  \eta^{Z_b} g_s(\mususy)^{\delta_{i1}(\delta_{j1}-1)} \left[\,X^{ab}_{ij}
  \, {\rm C}_j^{(1)}(\mususy) \right.\nonumber \\
&&\left. + \left(\delta
    X_{ij}^{ab}/\alpha_s(\mususy) + W_{ij}^{ab} + \eta^{-1}
    R_{ij}^{ab} \right) {\rm C}_j^{(0)}(\mususy) \right].
  \label{eq:magic64NLO1}
\end{eqnarray}
The relevant entries of the NLO magic numbers $\delta X_{ij}^{ab}$ ,
$W_{ij}^{ab}$ and $R_{ij}^{ab}$ are given  in Table \ref{tab:magicNLO}.
As expected, the evolution of $\vec{\rm C}^{(1)} (\mususy)$ is
dictated by the magic numbers derived in the LO case
(see eq.(\ref{eq:magic64})).

\subsection{NLO Wilson coefficients}
The computation of the matching conditions at the NLO level can be
divided in two parts: the matching conditions for the helicity flip
operators, $O^q_e, \, O^q_c$, and those for the Weinberg operator
$O_\gi$. Concerning the latter, the light quarks give a vanishing
contribution so that the only relevant contribution is due the top
quark that has no associated effective theory to subtract and
therefore no infrared (IR) divergent terms to deal with. Instead, for
helicity flip operators the computation of the NLO matching condition
is, in general, a very complicated task.  At the moment, even for a
process like $b \to s \gamma$, that has been investigated in great
detail in the last ten years, we have not yet obtained in the MSSM the
complete NLO matching conditions but only partial results are
available \cite{CDGG,DGG}.

We begin discussing the NLO matching conditions for $O_e,\, O_c$.
In the actual computation two strategies are
at hand. One can match matrix elements of operators belonging to a basis, like
the one in eq.(\ref{eq:ourops}), obtained enforcing the equation of motion,
a procedure that however requires, in general, an asymptotic expansion of
the relevant diagrams in the external momenta. Alternatively, one can use a
larger off-shell basis and perform the matching on the  off-shell  matrix
elements. In this case, one can use the freedom of the off-shell status
to choose a suitable kinematical configuration such that the relevant
Feynman diagrams can be evaluated using ordinary Taylor expansions in the
external momenta. The latter strategy, applied in ref.~\cite{CDGG1} to
the NLO matching conditions of the $\Delta B=1$
magnetic  and chromomagnetic operators, has been   employed by us in the
present work following closely ref.~\cite{CDGG1} to which we refer for
technical details.

The off-shell operator basis relevant for our calculation is obtained by
supplementing the basis in eq.(\ref{eq:ourops}) with the two
operators \cite{Simma}
\begin{eqnarray}
 O_m^q &=& -\frac{i}2   m_q \,\bar{q}  \gamma_5 q , \nn \\
 O_{p^2}^q &=& -\frac{i}2   \bar{q}\, \not\!\! D\not\!\! D \gamma_5 q
  \label{eq:offops}
\end{eqnarray}
where $D_\mu$ is the $SU(3)\times U(1)_Q$ covariant derivative. The relevant
terms in the Wilson coefficients are extracted via the use of the projector
\begin{eqnarray}
 P_{EDM}^\mu&=&\displaystyle{ \frac{1}{2\,q^6\,(n-2)} \left\{
q^2(n-2)(q^2-\not\!q\not\!p_1+m_q\not\!q)(2p_1-q)^\mu \right.}\nn \\
&&~~~+\left.m_q^2\left[(n-1)(4\not\!q\not\!p_1(2\,p_1-q)^\mu-4\,q^2p_1^\mu)+
2\,n\,q^2q^\mu-2\,q^2\gamma^\mu\not\!q\right]\right\}\gamma_5
 \label{eq:proje}
\end{eqnarray}
assuming an off-shell kinematical configuration defined by
$p_1^2=p_2^2=0,~p_1 \cdot q=q^2/2$ where $p_1$ and $q$ are the
momentum carried by the incoming quark and the external boson, respectively,
and $q^2 \ll m_q^2$.
The projector works by contracting it with the amplitudes and taking the trace,
so that  in eq.(\ref{eq:proje}) $\mu$ is the index carried by the external 
boson while $n$ is the dimension of the space-time. The use
of an off-shell kinematical configuration induces in the result for the
``full'' theory the appearance of terms that behave like $1/m_q^2, \,
\ln m_q^2$ as $m_q \to 0$. These infrared terms are eliminated by corresponding
terms in the effective theory once the off-shell basis $(O_e^q,\,
O_c^q,\,O_m^q,\,O_{p^2}^q)$ is employed so that the Wilson coefficients are
free of infrared terms as they should.

To simplify the calculation we compute the NLO gluino contribution to
the matching conditions retaining only one source of $CP$ violation,
namely we keep only one power of $X_q$, discharging terms $X_q^n$ with
$n >1$. We also work in the limit of $m_{\tilde{q}_{ L}} =
m_{\tilde{q}_{ R}}\equiv m_{\tilde{q}}$ with $ m_{\tilde{q}}$ common to
all squark flavours and taking all quarks massless but the top one. Within
this framework the NLO gluino corrections can be written
as
\bea
 C_{e \tilde{g}}^{q(1)}(\mususy) &=& \frac{\alpha_{s}}{4\pi\, m_{\tilde{g}}^2}
\left\{
\IM \left( \frac{X_q}{m_{\tilde{g}}} \right)\, \left[
F_1\left(x_{\tilde{g}}\right) + 4 F_2\left(x_{\tilde{g}}\right) +
F_2\left(x_{\tilde{g}},x_{t}\right)+
\RE \left( \frac{\mt\, X_t}{m_{\tilde{g}}^2} \right) \,
N_1\left(x_{\tilde{g}},x_t\right) \,\right]
\right.\nn \\
&&\left.~~+
\IM \left( \frac{\mt\, X_t}{m_{\tilde{g}}^2} \right) \,
\RE \left( \frac{X_q}{m_{\tilde{g}}} \right) \,
N_2\left(x_{\tilde{g}},x_t\right) \, \right\},
\label{eq:gluinotwoe} \\
C_{c \tilde{g}}^{q(1)}(\mususy) &=&
\frac{g_s \,\alpha_{s}}{4\pi\, m_{\tilde{g}}^2}
\left\{
\IM \left( \frac{X_q}{m_{\tilde{g}}} \right)\, \left[
F_3\left(x_{\tilde{g}}\right) + 4 F_4\left(x_{\tilde{g}}\right) +
F_4\left(x_{\tilde{g}},x_{t}\right) +
\RE \left( \frac{\mt\, X_t}{m_{\tilde{g}}^2} \right) \,
N_3\left(x_{\tilde{g}},x_t\right) \,\right]
\right.\nn \\
&&\left.~~+
\IM \left( \frac{\mt\, X_t}{m_{\tilde{g}}^2} \right) \,
\RE \left( \frac{X_q}{m_{\tilde{g}}} \right) \,
N_4\left(x_{\tilde{g}},x_t\right) \, \right\}
\label{eq:gluinotwoc}
\eea where in the above equations the upper line represents the $CP$
violation induced by the left-right entry in the mass matrix of the
squark of type $q$ while the lower one the corresponding effect due to
the stops. We have further divide the former contribution into the the
part due to the quark and squark of type $q$, that of the other four
squarks and massless quarks (the first two terms), and that due to the
top and stops including the mixing (the last two terms). In
eqs.(\ref{eq:gluinotwoe}-\ref{eq:gluinotwoc}) $x_{\tilde{g}}
=m_{\tilde{g}}^2/m_{\tilde{q}}^2$, $x_{t} =m_{t}^2/m_{\tilde{q}}^2$, ,
where the gluino and squark masses are assumed as $\overline{DR}$
parameters, and \be F_i = G_i + \Delta_i \, \ln
\frac{\mususy^2}{m_{\tilde{q}}^2}\quad,\quad
\lim_{x_{t}\rightarrow0}F_i(x_{\tilde{g}},x_{t})=F_i(x_{\tilde{g}}).
 \label{rgeev}
 \ee
The explicit expressions of the functions $G_i$ , $\Delta_i$, $N_i$
are reported in the Appendix. Defining
\bea
\tilde{\Delta}_1 &=& \frac{\alpha_s}{4\pi\, m_{\tilde{g}}^2} \IM
\left( \frac{X_q}{m_{\tilde{g}}} \right) \left( \Delta_1+5\Delta_2
\right), \nn \\
\tilde{\Delta}_2 & = & \frac{g_s\,\alpha_s}{4\pi\, m_{\tilde{g}}^2} \IM
\left( \frac{X_q}{m_{\tilde{g}}} \right) \left(\Delta_3+5\Delta_4
\right) \nn
\eea
we have that  the coefficients of the
$\ln( \mususy^2/m_{\tilde{q}}^2)$ terms satisfy
\be
\tilde{\Delta}_i= \frac12 \left[\left(1+i\right)\,\beta^{SUSY}_0+
           \sum_{k=\tilde{g},\tilde{q}} \gamma_{m_k}^{(0)} m_k
           \frac{\partial}{\partial m_k}+
           \gamma_{X_q}^{(0)}\right] C^{(0)}_i + \frac12 \sum_{j=1}^2
           \gamma_{ji}^{(0)} C^{(0)}_j\,,
\label{eq:R.G.E}
\ee
with
$$
\beta^{SUSY}_0 = 3\,N-n_f \quad,\quad
\gamma_{m_{\tilde{g}}}^{(0)}=2\left(3\,N-n_f\right) \quad,\quad
\gamma_{m_{\tilde{q}}}^{(0)}=4\,C_F\, m_{\tilde{g}}^2/m_{\tilde{q}}^2
\quad,\quad \gamma_{X_q}^{(0)}=-2\,C_F
$$
guaranteeing the cancellation of the $\mususy$ dependence to ${\cal O}(\als^2)$
in eq.(\ref{eq:wilsonexp}).
We observe that the effect of the term in the square brackets
in eq.(\ref{eq:R.G.E}) is to shift the coupling and the mass parameters
appearing in $C^{(0)}_i$ from the scale $\mususy $ to $m_{\tilde{q}}$.

\begin{figure}[t]
  \begin{center}
  \includegraphics[height=0.7\textwidth,angle=270]{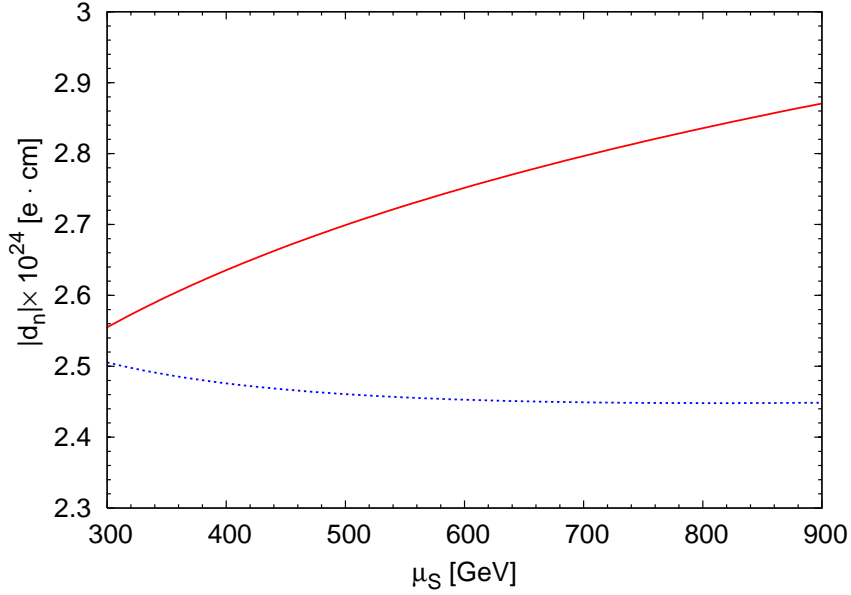}
 \caption{\itshape{
 The $\mu_S$ scale dependence of the gluino contribution to $d_n^e$
 at LO (upper curve) and NLO (lower curve).}}
    \label{fig:last}
  \end{center}
\end{figure}

As we anticipated in sec.~\ref{sec:LOuncertainties}, the inclusion of
the NLO matching for the gluino contributions reduces the perturbative
uncertainties down to a completely negligible level. Moreover the
inclusion of NLO corrections produces a non-negligible effect. In
fig.~\ref{fig:last} we plot the scale dependence of the gluino
contribution at the LO (upper line) and at the NLO (lower line) level.
As shown in the figure the inclusion of the NLO contribution greatly
reduces the scale dependence of the gluino contribution and lowers
$|d_n|$ of about 10\%.

Finally, for completeness, we consider also the Weinberg operator. At the
scale $\mususy$
two-loop diagrams where top and stops together with the gluino are exchanged
contribute to $C^{(1)}_\gi$. The relevant expressions can be gleaned from
ref.~\cite{Dai} obtaining
\be
C^{(1)}_{\gi \tilde{g}}  (\mususy) =
\frac{g_s\, \alpha_{s}}{4\pi\, m_{\tilde{g}}^2}
\IM \left(\frac{X_t}{ m_{\tilde{g}}} \right)
   \, H\left(x_{\tilde{g}},x_{t}\right),
\label{eq:wein}
\ee
where the function $H$ is found in the Appendix. However, when the evolution
down to a four-flavour theory is considered one has to take into account
also the shift in $C^{(1)}_{\gi}$ induced by the $O_c$ operator
at the $m_b$ threshold \cite{wise,braaten,Dai} or
\be
C^{(1)}_{\gi \tilde{g}}  (m_b^-) = C^{(1)}_{\gi \tilde{g}}  (m_b^+) +
\frac{\als(m_b)}{8\, \pi} C^{b(0)}_{c \tilde{g}}(m_b)~,
\label{eq:wmix}
\ee
where
\bea
C^{(1)}_{\gi \tilde{g}}  (m_b^+) &=&  {\eta_b}^{{39 \over 46}}
C^{(1)}_{\gi \tilde{g}}  (\mususy), \\
C^{b(0)}_{c \tilde{g}}(m_b) &=& {\eta_b}^{{5 \over 46}}
C^{b(0)}_{c \tilde{g}}(\mususy)
\eea
with $ \eta_b = \als (\mususy)/\als(m_b)$.

\section{Conclusions}
In this paper we have discussed the LO and NLO QCD corrections to the electric
dipole moment of the neutron in the MSSM. We pointed out the importance of
the mixing between the electric and chromoelectric operators that was always
neglected in previous analyses. Also we noticed that the QCD renormalization
factor of the dipole operator in absence of  mixing
should be less than 1 while its widely used estimate is $\eta^{ED} =1.53$
\cite{ALN}.

In the MSSM the prediction for the EDM can easily clash with the
experimental upper bound $|d_n|<6.3 10^{-26}$ e$\cdot$cm \cite{EDM} if the phases
in the mass parameters are
arbitrarily chosen. To avoid conflict with the experimental bound one can
consider  models with  approximate CP symmetries \cite{Dine} or flavour
off diagonal phases \cite{Abel} where small phases can be naturally obtained.
Another possibility is represented by a cancellation mechanism among different
contributions to the quark EDM which could allow O$(1)$ phases.
However we  noticed that, because of the mixing between the electric and
chromoelectric operators,  the neutralino contribution is always much more
suppressed than the gluino and chargino ones so that the cancellation
mechanism should actually mainly work  between the latter contributions that,
however, depend upon phases connected with apparently unrelated terms in the
MSSM Lagrangian.

Our results show that the NLO corrections we considered lower the prediction
of the EDM of about 10\% with respect to LO. Moreover the dependence on the
matching scale is drastically reduced. Clearly, a complete NLO analysis will
require, besides the gluino contribution we focused on, the other two
contributions, in particular the chargino one.

We have mainly focused on the perturbative aspects of the problem but, in order
to achieve a complete analysis of the SUSY constraints from the neutron EDM,
a lattice computation of the relevant matrix elements is mandatory.

Finally we notice that our analytic formulae for the magic numbers can
also be used for the evolutions of the Wilson coefficients in any
extension of the SM with new heavy particles, unless large four
fermion operators are generated at the matching scale.  This happens,
for example, in supersymmetric models with large $\tan\beta$
\cite{4fer}.

\section{Acknowledgments}
We thank A. Isidori for collaborating with us in the early stage of
this project. This work was supported in part by the EU network "The
quest for unification" under the contract MRTN-CT-2004-503369.

\begin{appendletterA}
\section*{Appendix}
In this appendix we give the explicit expressions of the one- and two-loop 
functions that appear in the Wilson coefficients.

The explicit expressions of the one-loop functions 
entering in eqs.~(\ref{eq:cegcn}-\ref{eq:neuc}) are given by
\bea 
\tilde{A}\left(x_1,x_2\right)&=&
-\frac{x_{1}x_{2}\left(x_{1}x_{2}+x_{1}+x_{2}-3\right)}
{2\left(x_{1}-1\right)^2\left(x_{2}-1\right)^2}-
\frac{x_{1}x_{2}}{\left(x_{1}-x_{2}\right)}
\left[\frac{x_{1}\ln\,x_{1}}{\left(x_{1}-1\right)^3}-\frac{x_{2}\,\ln\,x_{2}}
{\left(x_{2}-1\right)^3}\right]\nn\\
\tilde{B}\left(x_1,x_2\right)&=&
\frac{x_{1}x_{2}\left(3x_{1}x_{2}-x_{1}-x_{2}-1\right)}
{2\left(x_{1}-1\right)^2\left(x_{2}-1\right)^2}+
\frac{x_{1}x_{2}}{\left(x_{1}-x_{2}\right)}\left[\frac{x_{1}^2\ln\,x_{1}}
{\left(x_{1}-1\right)^3}-\frac{x_{2}^2\,\ln\,x_{2}}
{\left(x_{2}-1\right)^3}\right] \nn\\
\tilde{C}\left(x_1,x_2\right)&=&
\frac{x_{1}x_{2}\left(3x_{1}x_{2}+5x_{1}+5x_{2}-13\right)}
{3\left(x_{1}-1\right)^2\left(x_{2}-1\right)^2} \nn\\
&& -\frac{x_{1}x_{2}}{3\left(x_{1}-x_{2}\right)}
\left[\frac{x_{1}\left(x_{1}-9\right)\ln\,x_{1}}{\left(x_{1}-1\right)^3}-
\frac{x_{2}\left(x_{2}-9\right)\,\ln\,x_{2}}
{\left(x_{2}-1\right)^3}\right] \nn\\
A\left(x\right)&=&
\frac{x\left(x-3\right)}{2\left(x-1\right)^2}+
\frac{x\,\ln x}{\left(x-1\right)^3}\nn\\
B\left(x\right)&=&
\frac{x\left(x+1\right)}{2\left(x-1\right)^2}-
\frac{x^2\,\ln x}{\left(x-1\right)^3}~.
\eea
In the case of equal masses the functions $\tilde{A},\, \tilde{B},\, \tilde{C}$
reduce to
\bea
\tilde{A}\left(x\right)&=&
-\frac{x^2\,\left(x+5\right)}{2\,\left(x-1\right)^3}+
\frac{x^2\,\left(2x+1\right)\,\ln x}{\left(x-1\right)^4}\nn\\
\tilde{B}\left(x\right)&=&
\frac{x^2\,\left(5x+1\right)}{2\,\left(x-1\right)^3}-
\frac{x^3\,\left(x+2\right)\,\ln x}{\left(x-1\right)^4}\nn\\
\tilde{C}\left(x\right)&=&
\frac{2\,x^2\,\left(x+11\right)}{3\,\left(x-1\right)^3}+
\frac{x^2\,\left(x^2-16x-9\right)\,\ln x}{3\left(x-1\right)^4}~.
\eea

We list now the two-loop functions that appear in
eqs.~(\ref{eq:gluinotwoe}-\ref{eq:gluinotwoc}). To simplify the
expressions, we perform an expansion in the top mass reporting only
the first term in $x_t$. In particular, recalling eq.~(\ref{rgeev}),
we write
\bea
F_2 \left(x_{\tilde{g}},x_{t}\right) &=& G_2 \left(x_{\tilde{g}},x_{t}\right)
   + \Delta_2 \left(x_{\tilde{g}} \right) \,
  \ln \frac{\mususy^2}{m_{\tilde{q}}^2} \nn \\
  & \simeq &
G_2 \left(x_{\tilde{g}} \right) + x_t \, G_2^t
\left(x_{\tilde{g}} \right)
+ \Delta_2 \left(x_{\tilde{g}} \right)\,
\ln \frac{\mususy^2}{m_{\tilde{q}}^2} \nn \\
F_4 \left(x_{\tilde{g}},x_{t}\right) &=& G_4 \left(x_{\tilde{g}},x_{t}\right)
   + \Delta_4 \left(x_{\tilde{g}} \right)\,
\ln \frac{\mususy^2}{m_{\tilde{q}}^2} \nn \\
  & \simeq &
G_4 \left(x_{\tilde{g}} \right) + x_t \, G_4^t
\left(x_{\tilde{g}} \right) + x_t \ln x_t \, S_4 \left(x_{\tilde{g}} \right)
+ \Delta_4 \left(x_{\tilde{g}} \right)\,
\ln \frac{\mususy^2}{m_{\tilde{q}}^2} \nn \\
N_i \left(x_{\tilde{g}},x_{t}\right) &\simeq& \sqrt{x_t} \, N_i^t
\left(x_{\tilde{g}} \right) + \sqrt{x_t} \ln x_t \, R_i^t
\left(x_{\tilde{g}} \right)~. \nn
\eea
We find
\begin{eqnarray}
G_1 \left(x\right)&=&
     \frac{8 x^2 \left(51x^3+413x^2-1473x-251\right)}{27\left(x-1\right)^4}-
     \frac{16\left(8x^2+293x-13\right)}{27\left(x-1\right)^4}\,
              \Li_2\left(1-x\right) \nn \\
&& ~+\frac{8 x^2 \left(127x^4-1075x^3+480x^2+405x+27\right)\,\ln^2 x}
    {27\left(x-1\right)^6} \nn \\
&& ~-\frac{8x^2\left(48x^4+228x^3-1105x^2-548x+81\right)\,\ln x}
    {27\left(x-1\right)^5}
\label{eq:G1} \\
G_2 \left(x\right)&=&
     \frac{64 x^2\left(x^2+4x-2\right)}{3\left(x-1\right)^4}
    -\frac{16x^2\left(x+5\right)}{3\left(x-1\right)^3}\,\Li_2\left(1-x\right)
    -\frac{16x^3\left(x^2+4x-1\right)\ln x}{\left(x-1\right)^5} \\
G_3 \left(x\right)&=&
     -\frac{x^2\left(129x^3-2903x^2+1083x+21851\right)}{54\left(x-1\right)^4}-
     \frac{4\left(113x^2+281x+110\right)}{27\left(x-1\right)^4}\,
              \Li_2\left(1-x\right) \nn \\
&& ~+\frac{x^2\left(539x^4-2282x^3-6744x^2+7578x+621\right)\,\ln^2 x }
 {27\left(x-1\right)^6} \nn \\
&& ~+\frac{x^2\left(96x^4-5019x^3+13357x^2+10583x+1719\right)\,\ln x}
    {54\left(x-1\right)^5}
\label{eq:G3}\\
G_4 \left(x\right)&=&
    -\frac{x^2\left(17x^2-310x+101\right)}{3\left(x-1\right)^4}
    +\frac{2x^2\left(x-40\right)}{3\left(x-1\right)^3}\,\Li_2\left(1-x\right)
    +\frac{x^2\left(2x^3-67x^2+4x-3\right)\ln x}{\left(x-1\right)^5}
\label{eq:G4} \nn \\
G_2^t \left(x\right)&=&
     \frac{8 x^2\left(x^2+76x+175\right)}{9\left(x-1\right)^4}+
     \frac{32 x^2\left(x+3\right)}{\left(x-1\right)^5}\,\Li_2\left(1-x\right)-
     \frac{32x^3\left(2x+7\right)\ln x}{3\left(x-1\right)^5} \\
G_4^t \left(x\right)&=&
     -\frac{x^2\left(17x^2+230x-4387\right)}{18\left(x-1\right)^4}
    +\frac{2x^2\left(3x^2-8x+81\right)}{\left(x-1\right)^5}\,
     \Li_2\left(1-x\right) \nn \\
&& ~+\frac{x^2\left(53x^2-305x+18\right)\ln x}{3\left(x-1\right)^5}\ln x
\label{eq:G4t}\\
S_4 \left(x\right)&=&
     -\frac{x^2\left(x^2+10x+1\right)}{\left(x-1\right)^4}
    +\frac{6x^3\left(x+1\right)\ln x}{\left(x-1\right)^5}
\label{eq:S4}\\
N_1^t \left(x\right)&=&
     -\frac{4x^\frac{5}{2}\left(x^3+8x^2+173x+34\right)}{9\left(x-1\right)^4}
    -\frac{32x^\frac{7}{2}\left(x+2\right)\ln x}{3\left(x-1\right)^5} \nn \\
&& ~-\frac{16x^\frac{5}{2}\left(5x^2+17x+2\right)}{3\left(x-1\right)^5}
     \Li_2\left(1-x\right)
\label{eq:N1t}\\
N_2^t \left(x\right)&=&
     -\frac{4x^\frac{5}{2}\left(x^2+x+34\right)}{9\left(x-1\right)^3}
    -\frac{16x^\frac{5}{2}\left(x+2\right)}{3\left(x-1\right)^4}
     \Li_2\left(1-x\right)
\label{eq:N2}\\
N_3^t  \left(x\right)&=&
     -\frac{x^\frac{5}{2}\left(11x^3-158x^2+1225x+542\right)}
     {18\left(x-1\right)^4}+\frac{2x^\frac{5}{2}\left(2x^2-86x-9\right)\ln x}
     {3\left(x-1\right)^5} \nn \\
&& ~+\frac{2x^\frac{5}{2}\left(5x^2-181x-52\right)}{3\left(x-1\right)^5}
     \Li_2\left(1-x\right)
\label{eq:N3}\\
N_4^t  \left(x\right)&=&
     -\frac{x^\frac{5}{2}\left(11x^3-114x^2+753x-758\right)}
     {18\left(x-1\right)^4}-\frac{6x^\frac{5}{2}\left(6x-1\right)\ln x}
     {\left(x-1\right)^5} \nn \\
&& ~+\frac{2x^\frac{5}{2}\left(x^2-89x+52\right)}{3\left(x-1\right)^5}
     \Li_2\left(1-x\right)
\label{eq:N4}\\
R_3^t  \left(x\right)&=&
     -\frac{x^\frac{7}{2}\left(x^2-8x-17\right)}{2\left(x-1\right)^4}
    -\frac{3x^\frac{7}{2}\left(3x+1\right)\ln x}{\left(x-1\right)^5}
\label{eq:R3}\\
R_4^t  \left(x\right)&=&
     -\frac{x^\frac{7}{2}\left(x^2-8x-17\right)}{2\left(x-1\right)^4}
    -\frac{3x^\frac{7}{2}\left(3x+1\right)\ln x}{\left(x-1\right)^5}
\label{eq:R4}\\
\Delta_1 \left(x\right) &=& \frac{16 x^2 \left( 8x^3+13x^2-176x-37 \right)}
                            {9\left(x-1\right)^4} -
\frac{16x^2\left(29x^3-97x^2-115x-9\right)\ln x}{9\left(x-1\right)^5}
\label{eq:Delta1}\\
\Delta_2 \left(x\right) &=& \frac{8 x^2\left(7x^2+16x+1\right)}
                            {3\left(x-1\right)^4} -
\frac{16x^3\left(x^2+7x+4\right)\ln x}{3\left(x-1\right)^5}
\label{eq:Delta2}\\
\Delta_3 \left(x\right) &=& -\frac{8 x^2\left(2x^3-83x^2+268x+197\right)}
                            {9\left(x-1\right)^4} -
\frac{4x^2\left(59x^3-67x^2-643x-117\right)\ln x}{9\left(x-1\right)^5}
\label{eq:Delta3}\\
\Delta_4 \left(x\right) &=& \frac{2x^2\left(x^2+64x+31\right)}
                            {3\left(x-1\right)^4} +
\frac{2x^2\left(x^3-29x^2-59x-9\right)\ln x}{3\left(x-1\right)^5}
\label{eq:Delta4}
\end{eqnarray}
where ${\rm Li}_2(z) = -\int_0^z {\rm d}t \left[\ln(1-t)/t\right]$
is the dilogarithm function.

We observe that we have to take care of the fact that
the naive dimensional regularization (NDR) we used violate
supersymmetry, because the gauge boson and gaugino interactions with matter 
are different at one loop.
Supersymmetric Ward identity can be restored with an appropriate shift in the
gluino-squark-quark coupling and with a shift of the gluino mass
\cite{CDGG,Martin}.
Explicitly, the coupling and the gluino mass in the lowest order formula
(eqs.~(\ref{eq:gluinoe},\ref{eq:gluinoc})), must be replaced with
\bea
g_s & \rightarrow & g_s(1+\frac{\alpha_s}{4\pi}\frac{4}{3}) \\
m_{\tilde{g}} & \rightarrow & m_{\tilde{g}}(1+\frac{\alpha_s}{4\pi}3)~.
\label{eq:shift}
\eea

Finally, the function $H$ entering in the Weinberg operator
(eq.~(\ref{eq:wein})) is given by
\bea
H\left(x_{\tilde{g}},x_{t}\right) &\simeq&
H\left(x_{\tilde{g}}\right)+ x_t \, H_1^t\left(x_{\tilde{g}}\right)+
x_t \ln x_t \, H_2^t \left(x_{\tilde{g}} \right) \nn \\
H\left(x\right) &=&
   \frac{x^2\left(x+11\right)}{3\left(x-1\right)^3}+
   \frac{x^2\left(x^2-16x-9\right)\ln x}{6\left(x-1\right)^4} \nn \\
H_1^t\left(x\right) &=&
\frac{x\left(5x^3+265x^2+455x+27\right)}{6\left(x-1\right)^5}+
\frac{2x^2\left(4x^3-93x^2-258x-81\right)\ln x}
{3\left(x-1\right)^6} \nn \\
&& ~+\frac{2x^2\left(x^3-12x^2-51x-18\right)}{\left(x-1\right)^6}
 \Li_2\left(1-x\right) \nn \\
H_2^t\left(x\right) &=&
-\frac{x\left(11x^3-223x^2-259x-9\right)}{6\left(x-1\right)^5}+
\frac{x^2\left(x^3-12x^2-51x-18\right)\ln x}{\left(x-1\right)^6} ~~.
  \label{eq:H}
\eea
\end{appendletterA}

\end{document}